\documentclass[aps,prb,twocolumn,superscriptaddress,10pt]{revtex4-1}

\usepackage{graphicx}
\usepackage{amsmath}
\usepackage{amssymb}
\usepackage{amsfonts}
\usepackage{bm}
\usepackage{bbm}
\usepackage[mathcal]{euscript}

\usepackage{color}

\begin{document}
\title{Theory of local electric polarization and its relation to
internal strain: impact on the polarization potential and
electronic properties of group-III nitrides}

\author{Miguel~A. Caro}
\email{mcaroba@gmail.com}
\affiliation{Photonics Theory Group,
Tyndall National Institute, Dyke Parade, Cork, Ireland}
\affiliation{Department of Physics, University College Cork, Cork}
\author{Stefan Schulz}
\affiliation{Photonics Theory Group,
Tyndall National Institute, Dyke Parade, Cork, Ireland}
\author{Eoin~P. O'Reilly}
\affiliation{Photonics Theory Group,
Tyndall National Institute, Dyke Parade, Cork, Ireland}
\affiliation{Department of Physics, University College Cork, Cork}

\newcommand{\eq}[1]{Eq.~(\ref{#1})}

\begin{abstract}
We present a theory of local electric polarization in crystalline solids
and apply it to study the case of wurtzite group-III nitrides. We show that
a local value of the electric polarization, evaluated
at the atomic sites, can be cast in terms of a
summation over nearest-neighbor distances and Born effective charges.
Within this model, the local polarization shows a direct relation
to internal strain and can be expressed in terms of internal strain
parameters. The predictions of the present theory
show excellent agreement with a
formal Berry phase calculation for random distortions of a test-case
CuPt-like InGaN alloy and InGaN supercells with randomly placed cations.
While the present level of theory is appropriate
for highly ionic compounds, such as III-N materials,
we show that a more complex
model is needed for less ionic materials, such as GaAs,
in which the strain dependence of Born effective charges has to be
taken into account.
Moreover, we provide \textit{ab initio} parameters for
GaN, InN and AlN, including
hybrid functional values for the piezoelectric coefficients and the
spontaneous polarization, which we use to accurately implement the local
theory expressions.
In order to calculate the local polarization potential, we also present
a point dipole method.
This method overcomes several limitations related to discretization
and resolution
which arise when obtaining the local potential by solving Poisson's equation
on an atomic grid.
Finally, we perform tight-binding supercell calculations
to assess the impact of the local polarization potential arising
from alloy fluctuations on the electronic properties of InGaN alloys.
In particular,
we find that the large upward bowing with composition of the InGaN valence band
edge is strongly influenced by local polarization effects. Furthermore,
our analysis allows us to extract composition-dependent bowing parameters
for the energy gap and valence and conduction band edges.
\end{abstract}

\date{\today}

\maketitle

\section{Introduction}

Electrostatic built-in fields arising from discontinuities of the electric
polarization vector significantly modify the electronic and optical properties
of semiconductor nanostructures.~\cite{ambacher_2002,bester_2006,
beya_2011,caro_2012}
Of particular interest are systems such as GaAs-based quantum dots (QDs),
whose electronic and optical properties are affected by the symmetry of strain
and strain-induced piezoelectric fields.~\cite{grundmann_1995,schulz_2011}
The effect of built-in electrostatic fields
is even more dramatic in III-N-based heterostructures, where the large
piezoelectric response together with the intrinsic spontaneous
polarization give rise to built-in electrostatic
fields far exceeding those encountered for other
III-V materials.~\cite{takeuchi_1997,
ambacher_2002,kim_2007,lee_2010,caro_2011,schulz_2010,williams_2009}
Although these effects have been studied over the last two decades,
the possible role of the \textit{local} polarization
potential has only recently been considered.~\cite{caro_2012}

Theoretical studies that include a treatment of polarization
fields effectively treat the field
at a continuum level (even if the strain itself is
obtained from an atomistic calculation), with the polarization
assumed to have a smooth behavior with local strain and composition,
even in the case of alloys.
We have previously shown for InGaN alloys that a local value of
polarization can be obtained, observing large fluctuations in its value
at a microscopic scale.~\cite{caro_2012} In this paper we lay our theory
of local polarization on more solid ground, giving general equations
and providing a direct link with internal strain. We provide a
complete and consistent set of polarization-related \textit{ab initio}
parameters for the group-III nitrides, which are needed for the computation
of the local and macroscopic contributions to the total polarization.
In order to compute the electric potential arising from the local polarization,
we also present a ``point dipole'' method.

When computing the electronic properties of alloyed materials, it is of
vital importance that the supercell used allows to reproduce the different
configurations encountered in actual material samples. In practice, this
implies that the supercell must be sufficiently large. At present,
calculations for such large systems escape the reach of \textit{ab initio}
techniques, such as density functional theory (DFT).
Moreover, standard implementations of DFT fail to correctly describe
band gaps,~\cite{perdew_1983} and those implementations
that allow an accurate prediction
of this quantity, such as hybrid approaches,~\cite{seidl_1996}
are computationally much more expensive. On the other hand,
alternative semiempirical
electronic structure methods enable access to the electronic
properties of large systems for which first-principles approaches
cannot be realistically implemented. The tight-binding approximation
allows an accurate description of the electronic structure
in these cases, with
the advantage that polarization potentials and deformation potentials
can be included as on-site corrections to the Hamiltonian
matrix elements.~\cite{klimeck_2007,klimeck_2007b}
We therefore apply the tight-binding scheme in this work
in order to get insight into how the strong local polarization
effects influence the  electronic structure of InGaN alloys.

The paper is organized as follows. In Section~\ref{01} we introduce the
theoretical foundations of the present theory of local electric polarization
and discuss its degree of validity. In particular, we show in Section~\ref{07}
by comparing our local polarization results to DFT calculations
that the first-order level
of description presented here works remarkably
well in the case of group-III nitrides (relevant \textit{ab initio}
parameters for GaN, AlN and InN are given in Section~\ref{02}).
In Section~\ref{03} we present a point dipole method for the computation of
the local polarization potential on an atomic grid, and discuss practical
considerations regarding the implementation of the method.
Practical examples of the calculation of local polarization and local
polarization potential are given in Section~\ref{04} for polar and non polar
InGaN/GaN quantum wells (QWs).
In Section~\ref{05} we present a tight-binding (TB)
model for the calculation of the electronic structure in nitride systems,
and discuss how the local polarization potential affects the
band gap of InGaN. We then extract composition-dependent
bowing parameters for the band gap and for both the
conduction band (CB) and valence band (VB)
edges of InGaN alloys over the whole composition range in Section~\ref{62}.
Finally, we summarize our conclusions in Section~\ref{06}.

\section{Theory of local electric polarization}\label{01}

When treating a periodic crystal, it is usual to work in terms of the
dipole moment per unit volume, that is, the density of dipole moment,
or polarization. Crystals whose symmetry allows an
inversion centre cannot present a net
dipole moment.~\cite{nye_1985} For crystals without an inversion centre,
except point group 432,\cite{*[{432 is the Hermann-Mauguin symbol;
using the Schoenflies system, the equivalent point group is $O$ (orthorhombic
symmetry). While
point group 432 does not present \textit{linear} piezoelectricity, Grimmer has
shown that it is compatible with second-order
piezoelectricity. See }] [{.}] grimmer_2007}
certain deformations of the crystal lattice give origin to
net dipole moments, known as the \textit{piezoelectric} effect. In addition
to this, the subset of those crystals that present an anisotropic direction
in the lattice, called \textit{polar}, are compatible with
the existence of net dipoles even in the unstrained state, which is
referred to as \textit{spontaneous polarization}. The wurtzite (WZ)
crystal structure
belongs to the latter class and therefore WZ nitrides present both
piezoelectric and spontaneous polarization.~\cite{nye_1985}

The piezoelectric response of a material to strain is modeled, in the linear
regime,\footnote{One can also define a second-order piezoelectric tensor
to characterize piezoelectricity further away from equilibrium. See
Ref.~\onlinecite{grimmer_2007}} via the piezoelectric tensor $e_{ij}$:
\begin{align}
P_i^\text{pz} = \sum\limits_{j=1}^6 e_{ij} \epsilon_j,
\label{08}
\end{align}
where $P_i^\text{pz}$ are the components of the piezoelectric
polarization vector and $\epsilon_j$
are the strains, given in Voigt
notation.\footnote{Note that in Voigt notation,
$\epsilon_1 = \epsilon_{xx}$, $\epsilon_2 = \epsilon_{yy}$,
$\epsilon_3 = \epsilon_{zz}$, $\epsilon_4 = 2 \epsilon_{yz}$,
$\epsilon_5 = 2 \epsilon_{xz}$ and $\epsilon_6 = 2 \epsilon_{xy}$.}
The symmetry
of the crystal determines the non-zero elements of $e_{ij}$.
We shall see further on that, even for a bulk binary compound,
one can define a \textit{local} piezoelectric
tensor $e_{ij}^*$ whose \textit{average} over the unit cell
reduces to $e_{ij}$,
but that has in general more non-zero elements than $e_{ij}$. The total
polarization vector is given by
\begin{align}
P_i = P_i^\text{sp} + P_i^\text{pz},
\end{align}
where $P_i^\text{sp}$ are the components of the spontaneous polarization
vector, that will be present only if the crystal symmetry allows, as previously
discussed.

Calculating the polarization of a periodic crystal might seem at first
a trivial problem, with a possible intuitive definition being given by the
charge density of the unit cell. However, there is no way of unambiguously
defining the polarization vector using such a method, with an array of possible
values arising from different choices of origin.~\cite{resta_2007}
A rigorous frame for the computation of polarization in periodic solids was
not available until as recently as the 1990s. The main developments were
presented in the seminal papers by Vanderbilt and
King-Smith,~\cite{king-smith_1993,vanderbilt_1993} building up on an idea
originally suggested by Resta,~\cite{resta_1992} where the
foundations of the \textit{Berry-phase} theory of polarization,
or \textit{modern theory of polarization},~\cite{resta_1994} were laid.
This theory allows a calculation of the dipole moment of the unit
cell of a periodic insulating system,
which is well defined modulo $e \textbf{R}$
(where $e$ is the elementary charge and \textbf{R} is a lattice vector).
The latter ambiguity can be removed in different ways, such that a
meaningful value for the polarization can be
obtained.~\cite{king-smith_1993,vanderbilt_1993,bernardini_1997} However,
the obtainment of a position-dependent polarization vector, that varies
\textit{within} the unit cell in which the Berry phase is computed, is beyond
the reach of this technique. Nevertheless, for systems where composition and/or
strain change abruptly within the unit cell (e.g. random alloy InGaN QWs),
the question of whether a local value
of the polarization vector can be calculated becomes pertinent.

In the context of the Berry-phase technique,
only the average polarization
of the periodic unit cell as a whole can be calculated \textit{formally}.
In a general calculation, there may
not necessarily be an obvious or straightforward way to partition the
system into subsets for which the polarization can be easily computed
in separate calculations.
Any knowledge of how the polarization varies within the supercell
must therefore rely on a heuristic assumption.
This motivates to find a phenomenological solution to the problem,
to gain access to physical information which would not be
accessible otherwise. We show below that, within the present local
polarization formalism,
a position-dependent polarization, defined down to the unit volume
of an ensemble of nearest-neighbors,
yields results in good agreement with a \textit{formal}
Berry-phase calculation,
when extrapolated to calculate the average
polarization of the supercell.
This agreement provides strong support that the approach presented
here provides an accurate description of local polarization effects
in III-N heterostructures and alloys.

\subsection{Formal definition of the local polarization}\label{10}

As already discussed, the total macroscopic polarization has two components:
spontaneous and piezoelectric. Because the spontaneous polarization is a
reference state, establishing a local value for it formally
might prove rather non trivial: one would need to devise an adiabatic
transformation which keeps the system insulating while moving from
an equivalent centrosymmetric structure to the polar
crystal structure
that allows to evaluate the difference in polarization locally (at each atomic
site).~\cite{resta_2007} Therefore, to avoid this complexity,
we assume the spontaneous polarization for a given
binary compound to be position-independent and direct our attention towards
the piezoelectric polarization instead.

Our aim is a reformulation of \eq{08} that allows an evaluation
of the local and macroscopic contributions to the polarization separately.
For the sake of clarity and conciseness, we constrain ourselves to changes
in $P_i^\text{pz}$ that are linear in the strains. Future work will
extend our description to second-order piezoelectric polarization. As we will
see later on, the linear approximation breaks down quickly for some
III-Vs but is good up to moderate strain for the highly ionic
III-nitrides. In analogy to elasticity,~\cite{caro_2013}
we can generalize \eq{08} for arbitrary internal strains
as follows:
\begin{align}
P_i^\text{pz} = \sum\limits_{j=1}^6 e_{ij} \epsilon_j
+ \sum\limits_{\alpha=1}^{N_\text{atoms}} \sum\limits_{k=1}^3
\underbrace{\frac{\partial P_i^\text{pz}}{\partial
t^\alpha_k}}\limits_{e \mathcal{Z}^\alpha_{ik} / V} \left[ t^\alpha_k -
t^\alpha_{k,0} \left( \boldsymbol{\epsilon} \right) \right],
\label{11}
\end{align}
where  $N_\text{atoms}$
is the number of atoms in the unit cell, $t^\alpha_k$ is the $k^\text{th}$
component of the internal strain vector for atom $\alpha$,
$e$ is the elementary
charge, $V$ is the volume of the unit cell,
and $\mathcal{Z}^\alpha_{ik}$ is the
$ik$ component of the Born effective charge tensor~\cite{gonze_1997}
for atom $\alpha$.
$t^\alpha_{k,0} \left( \boldsymbol{\epsilon} \right)$ are the internal
strains that minimize the total energy of the crystal for any given strain
state $\boldsymbol{\epsilon}$.~\cite{caro_2013}
Although \eq{11} is general, because we are working in the linear
approximation we will assume that
the off-diagonal components of the Born effective charges are zero.
Equation~(\ref{11}) therefore reduces to
\begin{align}
P_i^\text{pz} = \sum\limits_{j=1}^6 e_{ij} \epsilon_j
+ \sum\limits_{\alpha=1}^{N_\text{atoms}} \frac{e \mathcal{Z}^\alpha_i}{V}
\left[ t^\alpha_i - t^\alpha_{i,0} \left( \boldsymbol{\epsilon} \right) \right],
\label{12}
\end{align}
where we have employed an implicit notation $\mathcal{Z}^\alpha_{i} \equiv
\mathcal{Z}^\alpha_{ii}$. Again, in the linear limit, the $t^\alpha_{i,0}$
are linear in $\boldsymbol{\epsilon}$ and we can write
\begin{align}
P_i^\text{pz} = \sum\limits_{j=1}^6 \underbrace{\left( e_{ij} -
\sum\limits_{\alpha=1}^{N_\text{atoms}} \frac{e \mathcal{Z}^\alpha_i}{V}
\frac{\partial t^\alpha_{i,0}}{\partial \epsilon_j} \right)}\limits_{
e_{ij}^{(0)}} \epsilon_j
+ \sum\limits_{\alpha=1}^{N_\text{atoms}} \frac{e \mathcal{Z}^\alpha_i}{V}
t^\alpha_i,
\label{13}
\end{align}
where $e_{ij}^{(0)}$ is the piezoelectric coefficient obtained
from a ``clamped-ion'' calculation,~\cite{bernardini_1997}
in which the ionic coordinates are not allowed to relax. Note that in
\eq{13}, the first term $e_{ij}^{(0)}$ is macroscopic, that is,
defined for the unit cell as a whole,
while the second one is evaluated locally.

Consider now that $V_0$ is the volume comprising an atomic site
and all of its nearest neighbors (in the context of the four-fold coordinated
ZB and WZ lattices this would correspond to each of the tetrahedra that
make up the crystal). We label the central atomic site 0 and each of its
nearest neighbors by $\alpha = 1,2,3, \dots, N_\text{coor}^0$. Then, the
relevant quantity in \eq{13} to be evaluated locally (at the atomic site 0) is
\begin{align}
P_{i,\text{local}}^\text{pz} (0) \equiv
\frac{e}{V_0} \left( \mathcal{Z}^0_i t^0_i +
\sum\limits_{\alpha=1}^{N_\text{coor}^0}
\frac{\mathcal{Z}^\alpha_i}{N_\text{coor}^\alpha} t^\alpha_i \right),
\label{14}
\end{align}
where $N_\text{coor}^\alpha$ is the number of nearest neighbors of
atom $\alpha$. By dividing the contribution of each of
the nearest neighbors $\mathcal{Z}^\alpha_i$ by their own number
of nearest neighbors $N_\text{coor}^\alpha$ we ensure
no double counting when extending the evaluation of \eq{14} to the whole
crystal.

The internal strains can be obtained in a relatively straightforward
manner for binary
compounds.~\cite{caro_2012c,caro_2013} However, for an irregular material,
such as an alloy, establishing a reference lattice structure with respect to
which the internal strains could be calculated would carry a
high degree of arbitrariness.
Furthermore, an \textit{exact} evaluation of \eq{14} would rely on
knowing the value of $\mathcal{Z}^\alpha_i$ for all the atoms present in
the crystal. For an irregular material, $\mathcal{Z}^\alpha_i$ would differ,
in general, for each atom, even (by a small amount)
for atoms of the same species.
Therefore, our choice is to deduce an
approximation to \eq{14} valid for a representative reference system (such as
a binary), and use that approximation to estimate the local polarization
in irregular systems.
We propose the following spherical approximation for the local environment
of the central atom (atomic site~0):
\begin{align}
\sum\limits_{\alpha=1}^{N_\text{coor}^0}
\frac{\mathcal{Z}^\alpha_i}{N_\text{coor}^\alpha} t^\alpha_i
\approx
- \frac{\mathcal{Z}^0_i}{N_\text{coor}^0}
\sum\limits_{\alpha=1}^{N_\text{coor}^0} t^\alpha_i.
\label{15}
\end{align}
The approximation given by \eq{15} would be exact if all the
nearest neighbors ($\alpha = 1,2,3,\ldots,
N_\text{coor}^0$) of atom~0 were piezoelectrically equivalent, that is,
if all of them have the same Born effective charges. This is the case
for binary ZB and WZ compounds. Further on, we will deal with how different
approximations work out for alloys.

\begin{figure}[b]
\centering
\includegraphics{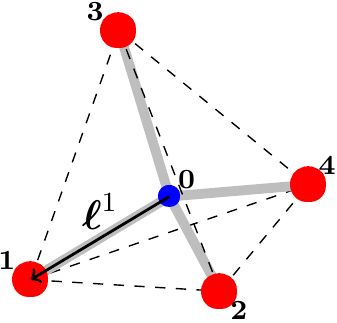}
\caption{(Color online)
First nearest-neighbor environment in a tetrahedrally bonded
crystal. The vector pointing from atom~0 (the central atom) towards
atom $\alpha$ is denoted $\boldsymbol{\ell}^\alpha$.}
\label{16}
\end{figure}

We can characterize the bonds between atom~0 and atoms $\alpha = 1,2,3,\ldots,
N_\text{coor}^0$ by a vector $\boldsymbol{\ell}^\alpha$ as indicated in
Fig.~\ref{16}. If $\boldsymbol{\ell}^\alpha_0$ is the bond vector of the
unstrained
case, we can write $\boldsymbol{\ell}^\alpha$ in terms of the macroscopic and
internal strains:
\begin{align}
\ell^\alpha_i = \sum\limits_{j=1}^3 \left( \delta_{ij} + \epsilon_{ij}
\right) \ell^\alpha_{j,0} + t_i^\alpha - t_i^0,
\label{17}
\end{align}
where $\epsilon_{ij}$ are the components of the strain tensor in Cartesian
notation and $\delta_{ij}$ is the Kronecker delta function.
With the approximation of \eq{15} and the definition given by
\eq{17} we rewrite \eq{13} as
\begin{widetext}
\begin{align}
P_i^\text{pz} = \sum\limits_{j=1}^6 e_{ij}^{(0)} \epsilon_j
- \frac{e}{V_0} \frac{\mathcal{Z}^0_i}{N_\text{coor}^0}
\left(
\underbrace{
\sum\limits_{\alpha=1}^{N_\text{coor}^0} \ell_i^\alpha}\limits_{\mu_i}
- \sum\limits_{j=1}^3 \left( \delta_{ij} + \epsilon_{ij} \right) \underbrace{
\sum\limits_{\alpha=1}^{N_\text{coor}^0} \ell_{j,0}^\alpha}\limits_{\mu_{j,0}}
\right),
\end{align}
where $\boldsymbol{\mu}$, defined as a summation over nearest-neighbor
distances, is the \textit{bond asymmetry parameter}.~\cite{caro_2012}
$\boldsymbol{\mu}_0$ is the bond asymmetry parameter of the unstrained system,
that would be zero for binary ZB materials and would have a
non-zero component along the polar axis $\mu_{3,0}$
for WZ materials.~\cite{caro_2012}

Finally, we write for the \textit{total} polarization at atomic site~0:
\begin{align}
P_i = \underbrace{\sum\limits_{j=1}^6 e_{ij}^{(0)}
\epsilon_j}\limits_\text{macroscopic} +
\underbrace{ P_i^\text{sp}
- \frac{e}{V_0} \frac{\mathcal{Z}^0_i}{N_\text{coor}^0}
\left( \mu_i - \sum\limits_{j=1}^3 \left( \delta_{ij} + \epsilon_{ij}
\right) \mu_{j,0} \right) }\limits_\text{local}.
\label{18}
\end{align}
\end{widetext}
Equation~(\ref{18}) is a central result of this paper, which separates the
contributions to the polarization arising from macroscopic effects, given by
the clamped-ion piezoelectric coefficient $e_{ij}^{(0)}$, and local effects,
dominated by internal strain.

\subsection{Validity of the model}

We have made a number of approximations in the previous section. Depending
on the nature of the compound at hand, each of them will have a different
impact on the results, and will limit the accuracy that can be achieved. 
These approximations are:
\begin{enumerate}
\item\label{19} We have assumed that $\textbf{P}_\text{sp}$
is constant throughout the crystal for binaries.
However, we have defined it as a \textit{local} quantity (this will
prove helpful when dealing with alloys).
\item\label{20} For the piezoelectric part, we have truncated
our description to first order in both macroscopic and internal strain.
\item\label{21} We have assumed that the off-diagonal terms of the Born
effective charge tensor are zero.
\item\label{22} We have performed a spherical approximation for the Born
effective charge of the nearest neighbors of the atom where the local
polarization is evaluated.
\end{enumerate}
As discussed in Section~\ref{10}, it is not trivial to establish whether
approximation~\ref{19} is good or not. It is possible to separate the
contributions to $\textbf{P}_\text{sp}$ into that arising from
the \textit{initial} bond asymmetry parameter $\boldsymbol{\mu}_0$
that we have defined previously
(which in WZ is related to the internal parameter $u$),
and the purely electronic contribution of the \textit{ideal} WZ
lattice.~\cite{caro_2012,
bernardini_2001,pal_2011} In this context, it is possible to assign a
local value for the initial bond asymmetry contribution,
which in the case of WZ would be equal in both cation and anion sites.
It seems therefore that assuming the electronic part to be also constant
between different atomic sites for the binaries might be reasonable.

\begin{figure}[t]
\includegraphics{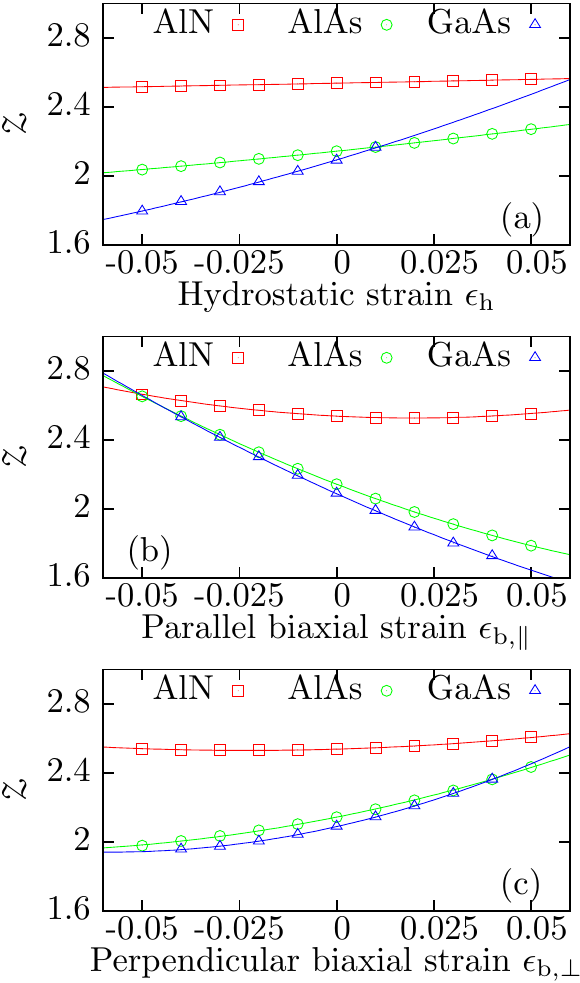}
\caption{(Color online)
Born effective charges of the corresponding cation for ZB AlN,
AlAs and GaAs, as a function of (a) hydrostatic and (b{\textendash}c)
biaxial strain. ``Parallel
biaxial strain'' means that the inequivalent
strain axis coincides with the axis along
which the Born effective charge is calculated, that is
$\epsilon_1 = \epsilon_2 = \epsilon_{\text{b},\parallel}$,
$\epsilon_3 = - 2 \epsilon_{\text{b},\parallel}$ and $\mathcal{Z}
\equiv \mathcal{Z}_3$. ``Perpendicular biaxial strain'' refers to the opposite
situation: $\epsilon_2 = \epsilon_3 = \epsilon_{\text{b},\perp}$,
$\epsilon_1 = - 2 \epsilon_{\text{b},\perp}$ and $\mathcal{Z}
\equiv \mathcal{Z}_3$. Open symbols are the results of LDA-DFT calculations
(see Section~\ref{07} for details) while solid lines are quadratic fits
to the data. The missing points for GaAs within this strain range
cannot be calculated
because the LDA predicts a conducting state which is not compatible
with the Berry-phase formalism (see discussion for InN
in Section~\ref{07}).~\cite{vanderbilt_1993,king-smith_1993}}
\label{23}
\end{figure}

Approximation~\ref{20} is indeed the main limitation to the model
introduced here,
but possibly the most straightforward one to overcome. The theory can
be extended to include second-order piezoelectric effects
at the expense of complicating
the formulas. We opt here to limit ourselves to a first-order description
to emphasize the conceptual implications of the theory. The linear limit
should be valid for highly ionic compounds, such as group-III nitrides,
as will be shown in the next section.
For the nitrides, although the second-order effects are large, the first-order
terms dominate up to strain values that are typically found in
realistic alloys and heterostructures (up to
5\%).~\cite{ambacher_2002,fiorentini_2002,caro_2011,prodhomme_2013}
However, for other III-V materials, second-order piezoelectric coefficients 
are relatively much larger compared to the linear ones. For instance,
for the Al
compounds AlP, AlAs and AlSb, Beya-Wakata \textit{et al}.~\cite{beya_2011}
found that the first-order piezoelectricity can practically be neglected
and second-order effects dominate even for small strains. For
GaAs the situation is intermediate and
the present level of theory should be accurate for small strains below
1 or 2~\%.
This complication is also present when computing the Born effective charges.
As we show in Fig.~\ref{23} for the hydrostatic and biaxial
strain dependence of $\mathcal{Z}$ (see figure caption and next section
for details of the calculation), the linear approximation
for the Born effective charge gets worse as one moves from highly ionic
AlN to the less ionic materials GaAs and AlAs.
Note that strain-dependent Born effective charges also have an impact
on the clamped-ion piezoelectric coefficient, as given by \eq{13}.
Therefore, a more complete and accurate treatment for general materials
should eventually include the dependence
of the Born effective charge $\mathcal{Z}_i$ on strain.
Note that within this linear model, the contributions
of clamped-ion terms and Born effective charges are
assumed linear in the strains in the formal derivation of the formulas.
However, the formalism does not \textit{impose} a linear dependence
of internal strain upon macroscopic strain when calculating the $\mu_i$:
this dependence is determined
by the specific theoretical framework used for the
computation of the atomic geometry of the system, e.g. DFT,
a valence force field, etc. In the case of nitrides, Prodhomme
\textit{et al.}~\cite{prodhomme_2013} have found relatively
large non-linear effects on binary and ternary compounds.
As will be shown in the next section, the present local
model succeeds at computing the polarization in nitride
ternaries because its main non-linear contribution arises
from non-linearities of local
internal strain \textit{itself}, including the effect
of disorder.

Approximation~\ref{21} is generally good, since for binary compounds the
off-diagonal components of the Born effective charge
are typically zero, and in any case the ratio $\mathcal{Z}_{ij} /
\mathcal{Z}_{ii} (i \neq j)$ is usually small.

The validity
of approximation~\ref{22} relies greatly on the specific crystalline
structure and whether the nearest
neighbors of the central atom where the polarization is being calculated
are equivalent (that is, have the same Born effective charge)
or not. For this reason, in the case of binary tetrahedrally
bonded compounds, where all the nearest neighbors for one given site
are of the same atomic
species, this approximation should be good for small strains. As
observed in Fig.~\ref{23} for biaxial strain, lattice distortions that
change the symmetry of the bonds have a large impact on the Born
effective charge for some compounds. Therefore, the validity of \eq{18}
would be limited for low ionicity and the more general form, \eq{13},
should be used. On the other hand, for ionic compounds such as nitrides,
\eq{18} retains its validity and offers an accurate description of the
local effects, as will be shown in Section~\ref{07}. In both cases (low
and high ionicity in tetrahedrally bonded binaries)
the approximation is exact for the linear piezoelectric limit (see
Section~\ref{24}).

\subsection{Testing the theory for group-III nitrides}\label{07}

As a first validation test and application of the theory, we have chosen
group-III nitrides.
The III nitrides are technologically important semiconductors
for a wide range of optoelectronic
applications.~\cite{nakamura_2000,krames_2007,mishra_2008}
The strong piezoelectric response of nitride compounds, together with the
existence of the spontaneous polarization, has a large impact on the
electromechanical properties of devices that incorporate them.
The large difference in bond lengths between the nitride binaries
leads to considerable local strains in these alloys, with measurable effects
such as large band gap bowings.~\cite{wu_2009,aschenbrenner_2010}
We have previously shown how these local strain fields affect the electric
polarization for InGaN alloys, retrieving the macroscopic limit with
the advantage of giving a description of the local effects at the same
time.~\cite{caro_2012} We have now presented in Section~\ref{01}
a refined and more general form of that model. In the following,
we will thoroughly apply this theory to test its validity for III-N materials.

\subsubsection{Parameters involved in the calculation of the local
polarization}\label{02}

\begin{table*}[t]
\caption{Parameters involved in the calculation of polarization-related
quantities for WZ group-III nitrides, obtained from DFT calculations
as explained throughout the
text. The HSE lattice parameters $a_0$, $c_0$, internal parameter $u_0$, and
internal strain parameters $\zeta_i$ are taken
from Ref.~\onlinecite{caro_2012c}.
The \textbf{k} grids are $6 \times 6 \times 4$ $\Gamma$-centered for a
four-atom hexagonal cell in
all cases except for the calculation of $e_{ij}$, $e_{ij}^{(0)}$,
$P^\text{sp}$ and
$\mathcal{Z}_i$ for InN in the LDA scheme.
For those quantities we use an orthorhombic-equivalent
16-atom supercell and the sampling in $k$ space is $4 \times 4 \times 4$,
following the standard Monkhorst-Pack scheme implemented in
\textsc{vasp}, which does not include $\Gamma$ in the integration
(see text for details).~\cite{ref_vasp} Note that in all cases, the
positive sign for $\mathcal{Z}_i$ implies a displacement of the cation
sublattice: the corresponding Born effective charge of the anions
is $-|\mathcal{Z}_i|$. $P^\text{sp}_\text{idWZ}$ is the spontaneous
polarization of the ideal WZ lattice (lattice parameters and internal
parameter extrapolated from the ZB phase).}
\begin{ruledtabular}
\begin{tabular}{l c c c c c c}
 & \multicolumn{2}{c}{AlN} & \multicolumn{2}{c}{GaN} & \multicolumn{2}{c}{InN}
\\
 & HSE & LDA & HSE & LDA & HSE & LDA \\
\hline \vspace{-0.8em} \\
$a_0$~(\AA)
& 3.103 & 3.092
& 3.180 & 3.154
& 3.542 & 3.507
\\
$c_0$~(\AA)
& 4.970 & 4.947
& 5.172 & 5.141
& 5.711 & 5.668
\vspace{0.2em} \\
$u_0$
& 0.3818 & 0.3820
& 0.3772 & 0.3765
& 0.3796 & 0.3787
\vspace{0.2em} \\
$\zeta_1$
& 0.138 & 0.145
& 0.156 & 0.168
& 0.193 & 0.204
\\
$\zeta_2$
& 0.086 & 0.091
& 0.083 & 0.089
& 0.107 & 0.112
\\
$\zeta_3$
& 0.191 & 0.200
& 0.159 & 0.168
& 0.218 & 0.226
\\
$\zeta_4$
& 0.199 & 0.224
& 0.201 & 0.210
& 0.337 & 0.339
\\
$\zeta_5$
& 0.143 & 0.140
& 0.141 & 0.148
& 0.107 & 0.118
\vspace{0.2em} \\
$e_{15}$~(C/m$^2$)
& -0.39 & -0.43
& -0.32 & -0.36
& -0.42 & -0.47
\\
$e_{31}$~(C/m$^2$)
& -0.63 & -0.69
& -0.44 & -0.49
& -0.58 & -0.63
\\
$e_{33}$~(C/m$^2$)
& 1.46 & 1.59
& 0.74 & 0.83
& 1.07 & 1.09
\vspace{0.2em} \\
$P_3^\text{sp}$~(C/m$^2$)
& -0.091 & -0.096
& -0.040 & -0.029
& -0.049 & -0.041
\\
$P_{3,\text{idWZ}}^\text{sp}$~(C/m$^2$)
& -0.031 & -0.033
& -0.019 & -0.016
& -0.019 & -0.016
\vspace{0.2em} \\
$e_{15}^{(0)}$~(C/m$^2$)
& 0.28 & 0.28
& 0.43 & 0.45
& 0.39 & 0.35
\\
$e_{31}^{(0)}$~(C/m$^2$)
& 0.26 & 0.25
& 0.40 & 0.41
& 0.37 & 0.38
\\
$e_{33}^{(0)}$~(C/m$^2$)
& -0.51 & -0.47
& -0.87 & -0.87
& -0.87 & -0.95
\vspace{0.2em} \\
$\mathcal{Z}_1 ( = \mathcal{Z}_2 )$
& 2.53 & 2.52
& 2.64 & 2.58
& 2.85 & 2.83
\\
$\mathcal{Z}_3$
& 2.68 & 2.67
& 2.77 & 2.72
& 3.02 & 3.00
\end{tabular}
\label{25}
\end{ruledtabular}
\end{table*}

The first step in setting up the theory
is to derive the necessary parameters for the WZ
III-N binaries GaN, AlN and InN: piezoelectric tensor $e_{ij}$, spontaneous
polarization $P^\text{sp}_i$, Born
effective charges $\mathcal{Z}_i$, lattice parameters $a_0$ and $c_0$,
internal parameter $u_0$, and internal strain parameters $\zeta_i$.
For our calculations
we have used the plane wave implementation of density functional theory (DFT)
available from the
\textsc{vasp} package,~\cite{ref_vasp,kresse_1996} within the projector
augmented-wave (PAW) method.~\cite{bloechl_1994,kresse_1999}
We perform calculations using both the local density approximation (LDA)
and the Heyd-Scuseria-Ernzerhof (HSE) screened-exchange hybrid
functional.~\cite{heyd_2003,heyd_2004,caro_2012c}
For the LDA calculations
we use \textsc{vasp}'s implementation of the Perdew-Zunger
parametrization,~\cite{perdew_1981} while the settings for the HSE functional
correspond to HSE06, with mixing parameter $\alpha = 0.25$ and screening
parameter $\mu = 0.2$. In all calculations the cutoff energy for plane
waves is 600~eV. All the quantities involving a calculation
of the polarization have been obtained using Martijn Marsman's
implementation of the Berry phase technique~\cite{vanderbilt_1993}
available in \textsc{vasp}.
We use HSE to obtain high quality parameters for the binaries
and LDA to perform test calculations for larger supercells and for statistical
evaluation of the accuracy of the theory.
In our experience, LDA-DFT gives a good
description of elastic properties and internal strain, while at the same
time being computationally affordable. Also, LDA-DFT seems to give
results in better agreement with experiments than generalized-gradient
approximations (GGAs) for the calculated electric polarization,
at least for III-V compounds.~\cite{beya_2011} The more computationally demanding
HSE functional, on the other hand, reduces
the band gap problem existent in standard Kohn-Sham DFT,~\cite{henderson_2011}
that potentially leads to a conducting phase being incorrectly
predicted for narrow gap semiconductors, such as InN. HSE also provides
lattice parameters and elastic properties in better agreement with
experiment.~\cite{caro_2012c}

The calculated structural and polarization-related
parameters of the III-N binaries are summarized in Table~\ref{25}.
In the context of the Berry phase approach,
a meaningful value for the polarization can only be calculated if the system
remains insulating.~\cite{vanderbilt_1993,king-smith_1993,resta_1994} As
already discussed, in the case of the III-N compounds
this is not a problem for the HSE
functional, which predicts a positive gap.~\cite{yan_2009} Using the LDA,
AlN and GaN are predicted to have (underestimated) positive gaps. However,
our settings lead to the prediction of a band crossing at the $\Gamma$ point
for InN, and therefore an incorrect metallic phase that
renders the calculation of a meaningful value of the polarization uncertain.
Previous data have been given for InN by Fiorentini and
collaborators in a series of papers on the piezoelectric properties and
spontaneous polarization of group-III
nitrides.~\cite{bernardini_1997,bernardini_2001,zoroddu_2001} While their
LDA calculations obtain the correct insulating phase of
InN,\footnote{Private communication with V.
Fiorentini and D. Vanderbilt.} ours must rely
on a different approach. Because the band crossing occurs only at the
$\Gamma$ point and immediate surroundings, we skip this area in the
$k$-point integration by shifting the \textbf{k} mesh away from $\Gamma$.
The resulting LDA
values of the polarization-related quantities in Table~\ref{25}
show almost perfect agreement with Fiorentini \textit{et al}.'s
LDA data,~\cite{bernardini_2001,zoroddu_2001} although InN remains
technically a metal in our case. The good agreement with the HSE calculation
further supports that our LDA values should be correct.

It should be noted that our calculations yield a negative sign for $e_{15}$
in both the LDA and HSE schemes.
Initial measurements~\cite{muensit_1999} and
calculations~\cite{bernardini_2002b} reported a positive value for $e_{15}$,
as included in Vurgaftman and Meyer's widely cited review
paper.~\cite{vurgaftman_2003} Our value here is in line with more
recent studies and analyses which show that a negative value is required
both for agreement with experiment and for internal consistency
among the different piezoelectric
coefficients.~\cite{schulz_2009,schulz_2011,schulz_2012b,shimada_2006}
Very recent LDA calculations of second-order
polarization of III-nitrides and ZnO by Prodhomme
\textit{et al}.~\cite{prodhomme_2013}
show good agreement with our linear coefficients of Table~\ref{25}.
The agreement between HSE and LDA
highlights the fact that LDA provides reliable
values for the electric polarization provided that it also succeeds
at predicting reliable band gaps and structural parameters: the largest
discrepancies are for the spontaneous polarization of GaN and InN, which
are influenced by the discrepancy between HSE and LDA for the calculated
value of $u_0$.

\subsubsection{Local piezoelectric tensor}\label{24}

We have previously obtained the
relation between macroscopic and internal
strain for the WZ lattice
and provided the definition of the five WZ internal strain parameters
$\zeta_i$ in Ref.~\onlinecite{caro_2012c}.
To obtain the relation between piezoelectric coefficients $e_{ij}$ and internal
strain parameters $\zeta_i$, one can apply \eq{13} to the internal strain
vectors for the WZ geometry.
The results can conveniently be expressed in the following compact form:
\begin{align}
e_{15} = e_{15}^{(0)} - \frac{2 e \mathcal{Z}_1}{\sqrt{3} {a_0}^2} \zeta_1,
\nonumber
\\
e_{31} = e_{31}^{(0)} - \frac{4 e \mathcal{Z}_3}{\sqrt{3} {a_0}^2} \zeta_2,
\nonumber
\\
e_{33} = e_{33}^{(0)} + \frac{4 e \mathcal{Z}_3}{\sqrt{3} {a_0}^2} \zeta_3.
\label{26}
\end{align}
We have incorporated in \eq{26} none of the assumptions leading to \eq{18}.
Therefore, \eq{26} is an exact result for WZ crystals in a linear piezoelectric
model.
It is thus initially
surprising that $\zeta_4$ and $\zeta_5$, although breaking the
cell symmetry, do not appear in the expressions for the $e_{ij}$. The
reason for this will become clear when obtaining the $e_{ij}$ as
$\frac{\partial P_i}{\partial \epsilon_j}$, calculated from \eq{18}.
\begin{figure}[t]
\includegraphics{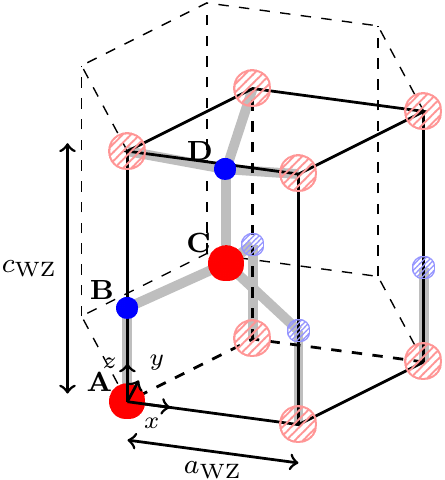}
\caption{(Color online) Standard four-atom WZ unit cell.
A and C are cations, B and D are anions.}
\label{27}
\end{figure}
Following
the convention of Fig.~\ref{27}, the \textit{local} piezoelectric tensor,
notated $e_{ij}^*$,
can be calculated at the atomic sites A and C, corresponding to the
two cations present in the unit cell, as the derivative of \eq{18} with
respect to the strains:
\begin{align}
e_{ij}^{*,X} = e_{ij}^{(0)} - \frac{e \mathcal{Z}_i^X}{\sqrt{3} {a_0}^2 c_0}
\left(
\frac{\partial \mu_i^X}{\partial \epsilon_j} - \sum\limits_{k=1}^3
\frac{\partial \epsilon_{ik}}{\partial \epsilon_j} \mu_{k,0}^X \right),
\label{28}
\end{align}
where $X$ indicates A or C.
For a WZ structure, the only non-zero component $\mu_{k,0}$
is $\mu_{3,0} = 4 (u_0 - 3/8 )
c_0$.~\cite{caro_2012} Expressing $\boldsymbol{\mu}$ in terms of macroscopic
strains, lattice parameters and internal strain parameters, each of the
non-zero components of $e_{ij}^*$ can be
obtained (an example
calculation for $e_{15}^{*,\text{A}}$ is given in Appendix~\ref{240}):
\begin{align}
& e_{15}^{*,\text{A}} = e_{15}^{*,\text{C}} = 
e_{15}^{(0)} - \frac{2 e \mathcal{Z}_1}{\sqrt{3} {a_0}^2} \zeta_1,
\nonumber
\\
& e_{16}^{*,\text{A}} = - e_{16}^{*,\text{C}} = 
\frac{\sqrt{3} e \mathcal{Z}_1}{2 a_0 c_0} \zeta_4 +
\frac{e \mathcal{Z}_1}{\sqrt{3} {a_0}^2} \zeta_5,
\nonumber
\\
& e_{21}^{*,\text{A}} = - e_{21}^{*,\text{C}} =  e_{16}^{*,\text{A}},
\nonumber
\\
& e_{22}^{*,\text{A}} = - e_{22}^{*,\text{C}} =  - e_{16}^{*,\text{A}},
\nonumber
\\
& e_{31}^{*,\text{A}} = e_{31}^{*,\text{C}} = 
e_{31}^{(0)} - \frac{4 e \mathcal{Z}_3}{\sqrt{3} {a_0}^2} \zeta_2,
\nonumber
\\
& e_{33}^{*,\text{A}} = e_{33}^{*,\text{C}} = 
e_{33}^{(0)} + \frac{4 e \mathcal{Z}_3}{\sqrt{3} {a_0}^2} \zeta_3.
\label{29}
\end{align}
That is, the expressions for $e_{15}$, $e_{31}$ and $e_{33}$ are retrieved
exactly, but additional piezoelectric components appear, that change
sign going from A to C.
To elucidate the effect of this on the
symmetry of the piezoelectric tensor, we write $e_{ij}^{*}$ in matrix
form:
\begin{align}
e_{ij}^{*,\text{A/C}} \equiv \left(
\begin{array}{c c c c c c}
0 & 0 & 0 & 0 & e_{15} & \pm e_{16}^*
\\
\pm e_{16}^* & \mp e_{16}^* & 0 & e_{15} & 0 & 0
\\
e_{31} & e_{31} & e_{33} & 0 & 0 & 0
\\
\end{array}
\right).
\label{30}
\end{align}
When averaging $e_{ij}^{*,\text{A}}$ and $e_{ij}^{*,\text{C}}$ within a
given unit cell,
one retrieves the WZ macroscopic limit:
\begin{align}
\frac{1}{2} \left( e_{ij}^{*,\text{A}} + e_{ij}^{*,\text{C}}
\right) \equiv \left(
\begin{array}{c c c c c c}
0 & 0 & 0 & 0 & e_{15} & 0
\\
0 & 0 & 0 & e_{15} & 0 & 0
\\
e_{31} & e_{31} & e_{33} & 0 & 0 & 0
\\
\end{array}
\right).
\label{31}
\end{align}
The anion sites B and D have the same expressions for $e_{15}$,
$e_{31}$ and $e_{33}$ and slightly different expressions for $e_{16}^{*}$:
\begin{align}
e_{16}^{*,\text{B}} = - e_{16}^{*,\text{D}} = &
- \frac{\sqrt{3} e \mathcal{Z}_1}{2 a_0 c_0} \zeta_4
- \frac{2 e \mathcal{Z}_1}{\sqrt{3} {a_0}^2} \zeta_5.
\end{align}
The macroscopic limit is of course also retrieved when averaging for the
anion sites. Note that the values of $e_{16}^*$ are comparable to those
of the macroscopic piezoelectric tensor coefficients. For instance, for GaN,
$|e_{16}^*|$ amounts to
0.79~C/m$^2$ and 1.13~C/m$^2$ for cation and anion sites, respectively.

Equation~(\ref{30}) is the (site-dependent) local piezoelectric tensor for
a WZ
lattice. It reflects the fact that there exist two sets of inequivalent
tetrahedra in a WZ lattice, and that the macroscopic strain affects the
nearest-neighbor environment of each of them
differently.~\cite{caro_2012c,caro_2013}
This is \textit{a priori} an unexpected result,
and implies that crystals that are non-polar and
non-piezoelectric \textit{on average} could nevertheless
present a local, perhaps measurable,
piezoelectric-like polarization.

Finally, note the similarity between the local
piezoelectric tensor of WZ and that of ZB in a (111)-oriented description
[Eq.~(27) of Ref.~\onlinecite{schulz_2011}]. This reflects that
(111)-oriented ZB systems
present a three-fold symmetry,~\cite{schulz_2011}
where all cation (anion) sites have an
equivalent environment, contrary to the WZ case, where there are two
inequivalent cation (anion) sites.~\cite{caro_2012c}

\subsubsection{Local polarization in InGaN alloys: strategies and testing}

\begin{figure}[t]
\includegraphics{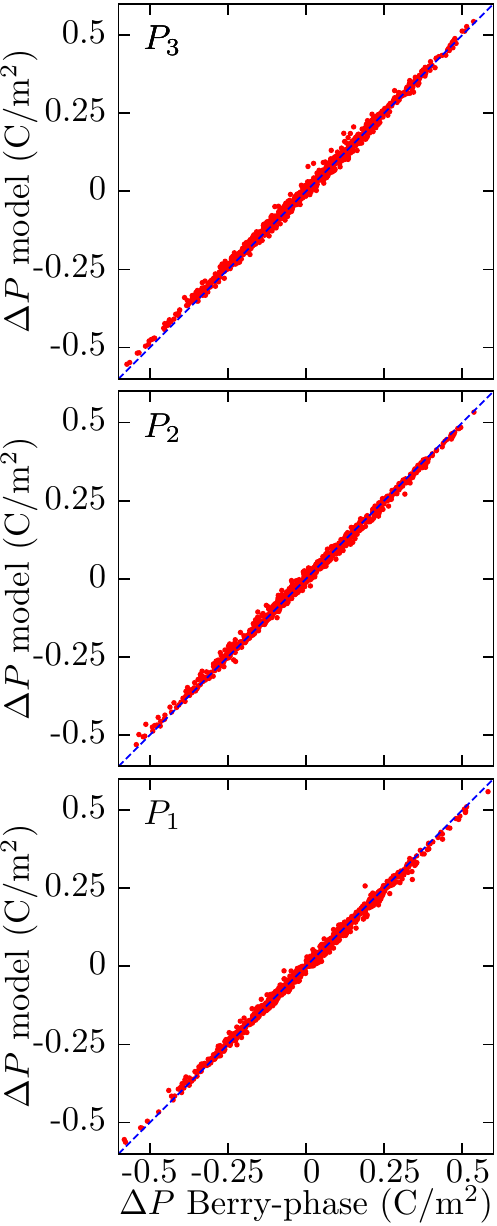}
\caption{(Color online)
Comparison between the polarization predicted by the present model
and a Berry-phase calculation for a large number (1000) of randomly
distorted CP-like InGaN cells (four-atom unit cell).
$\Delta P$ is the difference in polarization between the equilibrium
and distorted structures, where the lattice vectors are fixed
but the coordinates of each atom in the
unit cell are varied randomly up to $\pm 0.2 \text{~\AA}$
in each Cartesian direction.
The Berry-phase values are LDA-DFT results. The dashed line indicates
perfect agreement between the two methods, that is $\Delta P_\text{model}
= \Delta P_\text{Berry-phase}$. A few random distortions within the range
lead to a metallic phase being predicted by LDA, and
were left out of the comparison.}
\label{32}
\end{figure}

\begin{figure*}[t]
\begin{tabular}{c c c}
\textbf{32-atom supercells} & \hspace{0.1cm} & \textbf{128-atom supercells}
\vspace{0.4cm}
\\
\includegraphics{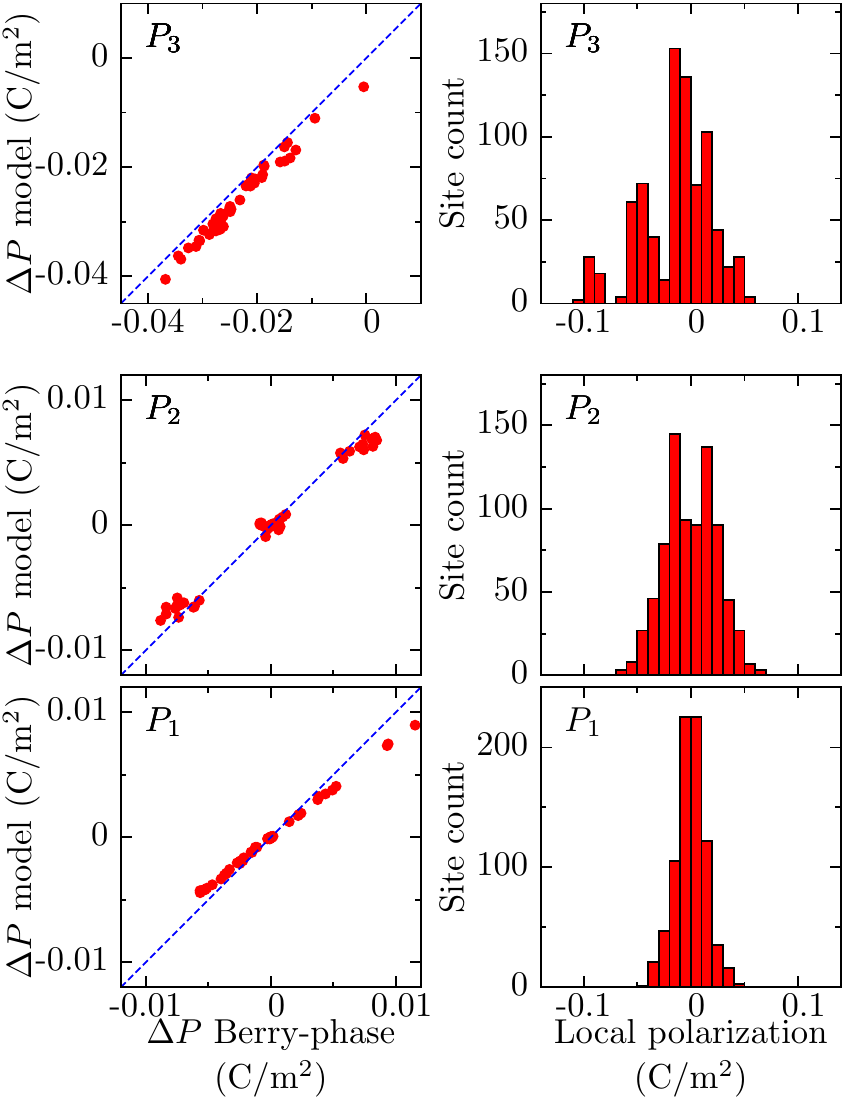} &
& \includegraphics{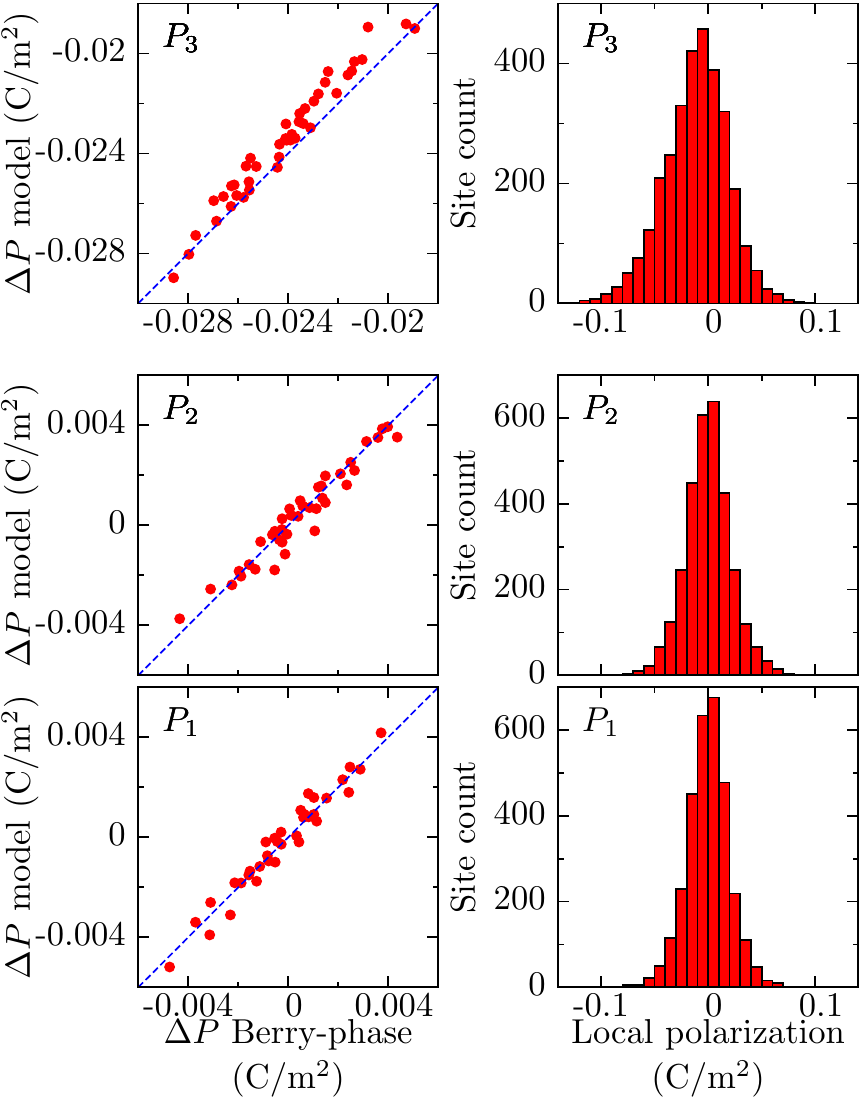}
\end{tabular}
\caption{(Color online)
Comparison between the spontaneous polarization calculated using
the present local model and the Berry-phase technique for a series
of In$_{0.5}$Ga$_{0.5}$N random supercells with 32 and 128 atoms. $\Delta P$
is the difference between the polarization of the supercells before and
after internal strain relaxation. The Berry-phase values and the
relaxed atomic positions are LDA-DFT results.
The dashed line indicates
perfect agreement between the two methods, that is $\Delta P_\text{model}
= \Delta P_\text{Berry-phase}$. The supercells are orthorhombic, with
the lattice vectors given by the average of the InN and GaN lattice
parameters. In terms of cation layers, the 32-atom supercells have
a $2 \times 4 \times 2$ arrangement while the 128-atom supercells are
$4 \times 4 \times 4$. The ``site count'' panels for each series refers
to the number of cation sites that registered a local polarization
value within the ranges shown (of width 0.01~C/m$^2$),
for the combined supercells.}
\label{33}
\end{figure*}

We have seen so far that for wurtzite
nitride binaries there is an exact correspondence
between local and macroscopic polarization that is retrieved
when averaging the local part over the unit cell. Although some solid-state
devices might operate employing binary compounds, the most interesting
applications of the nitrides arise through the use of their alloys
for controlled variation of properties (e.g. band gap tunability).

The main problem facing a local polarization
calculation for an alloy
is the increased complexity of the atomic environment of each of the
sites where the local polarization is to be evaluated. This is due to the
fact that the Born effective charges of all the atoms involved in the
calculation are affected by the interaction with all the other atoms
present in the crystal. In a periodic cell calculation this number would be
reduced to the number of atoms in the supercell. Since there is an
arbitrarily large number of possible configurations depending on alloy
composition and supercell size, establishing an exact correspondence
between local and macroscopic polarization in the fashion of Section~\ref{10}
then becomes virtually impossible. To overcome this limitation, we will assume
for the nitrides, and InGaN in particular, that the Born effective charge
of the cations in the alloy remains the same as for the binary, and that the
spherical approximation still holds.~\cite{caro_2012} We have devised two
tests in order to establish how good this approximation is.
First, we will use the smallest alloy cell, which is
a CuPt-like (CP-like) InGaN unit cell consisting only
of 4 atoms,~\cite{caro_2012,bernardini_2001} and will perform random
distortions of the atomic positions
within the unit cell. The result of the averaged local polarization, calculated
using \eq{18}, will be compared to the formal Berry-phase result.
Second, 32- and
128-atom In$_{0.5}$Ga$_{0.5}$N supercells will be considered and the cation
sites occupied randomly with
either a Ga or an In atom, with the only requirement
that the stoichiometric ratio of 1/1 be preserved (i.e. the nominal
composition of all cells is the same). The internal atomic
positions will then be allowed to relax by minimizing the supercell LDA-DFT
total energy, and the result of the averaged
local polarization will again be compared to that of a Berry-phase calculation.
The statistical treatment of both tests will reveal the validity of the
approximation for InGaN alloys.

The results of the first test are depicted in Fig.~\ref{32}. The figure
shows a comparison of the average polarization of the CP-like InGaN
cell calculated both within the present local polarization model and
with the Berry-phase technique.
We have performed random displacements of up to $\pm 0.2 \text{~\AA}$ (which
is equivalent to approximately 10\% of the equilibrium bond lengths) to
each of the Cartesian coordinates of each of the 4 atoms in the unit cell.
For the local polarization model, we have computed the local polarization
contributions at the Ga and In sites using \eq{18} and then obtained
its average for the whole cell.
Since only differences in polarization are meaningful within the
Berry-phase formalism,~\cite{king-smith_1993,vanderbilt_1993} we compare in
Fig.~\ref{32} the difference $\Delta P$ between the polarization of
the equilibrium CP-like InGaN structure and the distorted one.
As can be seen, the agreement between the two methods is excellent,
with all the data points lining up against the dashed line that corresponds
to perfect agreement $\Delta P_\text{model} = \Delta P_\text{Berry-phase}$.

Even more enlightening is the comparison between the present model
and the Berry-phase results depicted in Fig.~\ref{33} for random
In$_{0.5}$Ga$_{0.5}$N orthorhombic supercells. In that figure, $\Delta P$ is
the difference
between the polarization of the supercell before and after
internal strain relaxation. The supercells are constructed with either 32
or 128 atoms and the In and Ga atoms are placed randomly at the cation sites.
The lattice vectors of the supercells are kept fixed and chosen as the
average between the LDA values for the binaries. The ``site count'' panels
show the number of cation sites that present a particular
\textit{local} polarization value within different ranges, for the combined
supercells. We note two main features. The first observation is that
the local polarization model succeeds at very accurately predicting
the average supercell polarization even though the latter is calculated
from a sum over many local contributions whose values vary within limits
which are approximately
one order of magnitude higher. Second, our results show that the
average polarization is highly dependent on the specific atomic arrangement,
even for a large number of atoms. Bernardini and
Fiorentini~\cite{bernardini_2001} have previously calculated the spontaneous
polarization for the same material using a 32-atom special quasirandom
structure (SQS),~\cite{wei_1990} and have proposed that disorder plays only
a secondary role in the calculation of the polarization, both spontaneous
and piezoelectric.~\cite{bernardini_2001,bernardini_2002,fiorentini_2002,
ambacher_2002} We have found that this is indeed the case for the spontaneous
polarization of the supercells studied \textit{before} the optimization of the
atomic degrees of freedom: all the 128-atom configurations studied yielded
the same value of $\sim -0.009 \text{ C/m}^2$ within less than 0.001~C/m$^2$
of each other. However, our results suggest i) that a 32-atom supercell
might not be large enough to study the effect of disorder (see e.g. clustering
of calculated values for $P_2$ in Fig.~\ref{33}) and ii) that
internal strain relaxation introduces large corrections to the polarization
value, even for supercells containing as many as 128 atoms. Note, for instance,
that the average in-plane components of the polarization $P_1$ and $P_2$,
which are not symmetry-allowed for the binaries, do not vanish for the alloys
in the case of finite-size supercells.

All of these considerations not only support the validity of the local
model discussed here, but also highlight the need for one, in order to be
able to treat the effects of disorder and associated internal strain
accurately.

\section{Point dipole method for the calculation of the polarization
potential}\label{03}

When trying to calculate the local polarization potential
by solving Poisson's equation
\mbox{$\nabla \cdot (\varepsilon \nabla \phi ) = \nabla \cdot \textbf{P}$}
in the same atomic grid where
the polarization is given, one encounters two main difficulties.
The first arises from
the discretization of the medium, which is irregular given the
arrangement of the atoms in the strained crystal. The second, and most
important, is a problem of resolution: because Poisson's equation
needs to be solved in a finite difference or polynomial interpolation schemes,
and its solution
involves the calculation of several derivatives (see, for instance,
Ref.~\onlinecite{bester_2005}), approximate interpolations have to
be made and the effects of abrupt local discontinuities are lost in
the process. In order to compute the local polarization potential
and overcome these limitations,
we have previously used a point dipole model.~\cite{caro_2012}
Here we give the details of our model and extend it, as well as assess
its limitations and degree of
validity for calculations involving a position-dependent value of
the polarization.

The point dipole model is a solution to
the challenge of solving Poisson's equation on an atomic
grid where abrupt changes in the polarization vector occur.~\cite{caro_2012}
This is
achieved with a method that computes at any arbitrary position
the potential contribution due to each
dipole individually, without involving the interpolation of quantities
between neighboring grid sites that would lead to loss of resolution.
However, before the polarization potential can be obtained from the
point dipoles, a remapping of polarization density into dipole moment
on the system's grid has to be performed. The latter is dealt with
in Section~\ref{34}. The general solution for the polarization potential
arising from the ensemble of point dipoles is obtained in Section~\ref{35}
in an image dipole scheme, for a QW system (or layered structure, in general)
where a different arbitrary dielectric
constant is allowed for all three neighboring layers of material. The effect
of different levels of approximation for this general solution is also treated
in Appendix~\ref{241}.
In Section~\ref{36} we present a comparison
between the solution of Poisson's equation for a problem with an available
analytical solution and different levels of implementation of our method.
Further material complementary to this section, including computational
aspects, is given in Appendix~\ref{241}.

\subsection{From polarization to dipole moment}\label{34}

Before establishing the form of the potential due to a point dipole
ensemble, we focus
our attention on the transformation between polarization density
$\textbf{P}$,
which is the quantity usually calculated in strained crystals,
and dipole moment $\textbf{p}$,
which is the quantity involved in the equations that will be presented
in the next section.

The polarization $\textbf{P}$ can be
understood as a ``density of dipole moment''. Indeed, the total
dipole moment of
a finite size sample in which the polarization density is constant is simply
the product of $\textbf{P}$ and the volume of the sample.
Therefore, when dealing with constant polarization in a continuum-based
description, a standard cubic
discretization of the material, with step size $\Delta$,
is well suited to the
representation of $\textbf{P}$ as an ensemble of dipoles of magnitude
$\textbf{p} = \textbf{P} \Delta^3$ located at each of the mesh points.
However, our main interest is the representation of the material as an
ensemble of point dipoles in an atomistic scheme. For tetrahedrally bonded
compounds this involves the discretization in a mesh with either
cubic (zinc-blende)
or hexagonal (wurtzite) coordination, in the ideally undistorted lattice. After
strain is applied, the former grids will suffer a deviation from cubic
and hexagonal symmetries and the assignment of a finite volume to each
mesh point becomes cumbersome.

In the description of local polarization that we have previously
employed, the values of $\textbf{P}$ were given at the sites
of each of the cations
present in the crystal.~\cite{caro_2012}
The latter is a useful description, in the sense that the
representation of the whole crystal as a collection of deformed
tetrahedra can be done via the relative positioning of the
nearest neighbors: each cation and its four neighboring anions unambiguously
define each tetrahedron. Labeling the anions immediately surrounding a cation
as 1, 2, 3 and 4 (Fig.~\ref{16}),
we refer to the volume of the corresponding tetrahedron
as $V_\text{1234}$. If the positions of the anions are
$\textbf{r}_\text{1}$, $\textbf{r}_\text{2}$, $\textbf{r}_\text{3}$ and
$\textbf{r}_\text{4}$, then $V_\text{1234}$ is given by
\begin{align}
V_\text{1234} = 
\frac{1}{6} | \left( \textbf{r}_\text{1} - \textbf{r}_\text{4}
\right) \cdot \left[ \left( \textbf{r}_\text{2} - \textbf{r}_\text{4}
\right) \times \left( \textbf{r}_\text{3} - \textbf{r}_\text{4} \right)
\right] |.
\end{align}
However, it can be easily shown that $V_\text{1234}$ only accounts for
the volume of the tetrahedron itself and that a summation of the volumes
of all the tetrahedra contained
within a material sample would underestimate
the volume of the sample by exactly a factor of 6. Therefore, we
define the volume \textit{corresponding} to a tetrahedron as
\begin{align}
\tilde{V}_\text{1234} &= 6 V_\text{1234}.
\label{38}
\end{align}
Now, the value of the dipoles can be easily obtained once a map of the
polarization is available. For simplicity, we denote each grid point by
$i$ and the volume of the corresponding tetrahedron, as given by
Eq.~(\ref{38}), as $\tilde{V}_i$:
\begin{align}
\textbf{p}_i = \textbf{P}_i \tilde{V}_i,
\end{align}
with $\textbf{p}_i$ being located at the position $\textbf{r}_i$ of cation $i$.

Our choice for a cation-based description stems from convenience. In a
nitride alloy all the anions are nitrogen atoms and therefore
applying the spherical
approximation of \eq{15} (which is based on nearest neighbors only)
leads to one Born effective charge definition per cation atomic species: Ga, In
and Al for conventional III-N.
Using an anion-based description would lead, in the case of nitrides,
to defining 15 different
Born effective charges for N, which correspond to the 15 possible
combinations of Ga/In/Al atoms that can be nearest neighbors to N (e.g,
4 Ga, 3 Ga and 1 In, 2 Al and 2 In, etc.).

\subsection{Solution for materials with different dielectric
constant}\label{35}

Given the multipole expansion of a distribution of electric charge (see, for
example, Ref.~\onlinecite{jackson_1999}), the contribution to the
electrostatic potential $\phi_{\textbf{p}}$ calculated at
$\textbf{r}$ due to a point dipole $\textbf{p}$ is given by
\begin{align}
\phi_{\textbf{p}} \left( \textbf{r} \right) = \frac{1}{4 \pi \varepsilon_0
\varepsilon_r} \frac{\textbf{p} \cdot \left( \textbf{r} - \textbf{r}_\textbf{p}
\right)}{|\textbf{r} - \textbf{r}_\textbf{p}|^3},
\label{39}
\end{align}
where $\textbf{r}_\textbf{p}$ is the position
of the dipole $\textbf{p}$, $\varepsilon_0$
is the permittivity of the vacuum, and
$\varepsilon_r$ is the dielectric constant of the material.
Equation~(\ref{39}) is only valid when both the dipole $\textbf{p}$ at
$\textbf{r}_\textbf{p}$ and the point $\textbf{r}$ where the potential
is calculated are contained within an infinite (or big enough to neglect
surface effects) sample of a dielectric material with dielectric
constant $\varepsilon_r$. For the more general case in which there are
boundaries between materials with different dielectric constants,
e.g. a quantum well, it is appropriate to use the method of images
to obtain a form of Eq.~(\ref{39}) that accounts for the discontinuity
of $\varepsilon_r$ across the different interfaces. The details of the
method and the treatment for the case of up to three material layers with
different dielectric constants are given in Appendix~\ref{241}.

\subsection{Comparison to the solution of Poisson's equation for simple
structures}\label{36}

\begin{figure}[t]
\includegraphics{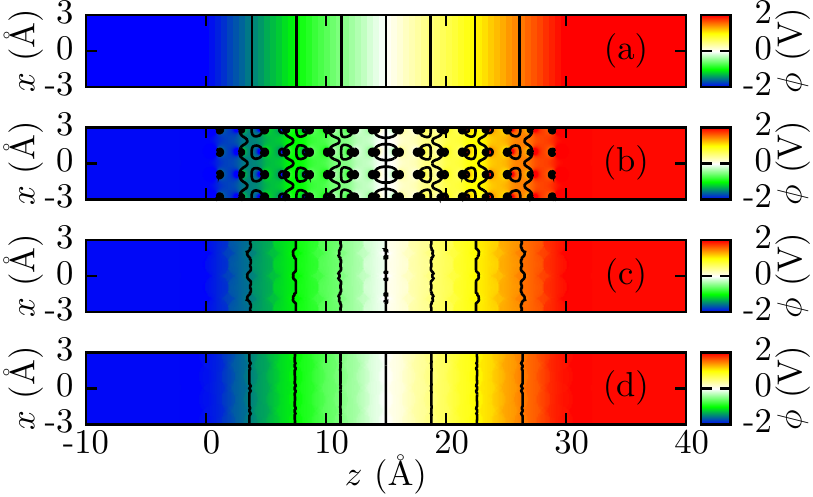}
\caption{(Color online) Potential obtained at different levels of
approximation for a QW of width
$h = 30 \text{ \AA}$ for which
$P_0 = 0.1 \text{ C/m}^2$ and $\varepsilon_r = 8.4$. In the barrier
$P = 0$ and $\varepsilon_r = 9.6$. (a) Analytic solution
[Eq.~(\ref{42})], (b) direct application of the present dipole method,
(c) dipole method with cutoff radius
$r_\text{cutoff} = 1 \text{ \AA}$ and (d)
dipole method with Gaussian smearing implementation,
$r_\text{smear}=1.5\text{ \AA}$ and $\sigma = 1\text{ \AA}$ (see
Appendix~\ref{241}).}
\label{43}
\end{figure}

\begin{figure*}[t]
Polar \hspace{5cm} Non polar\\
\includegraphics{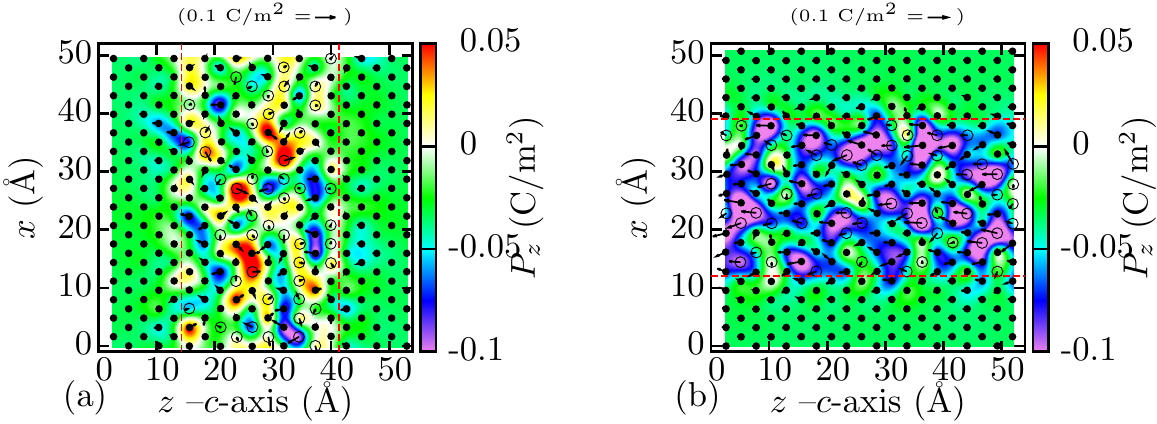}
\includegraphics{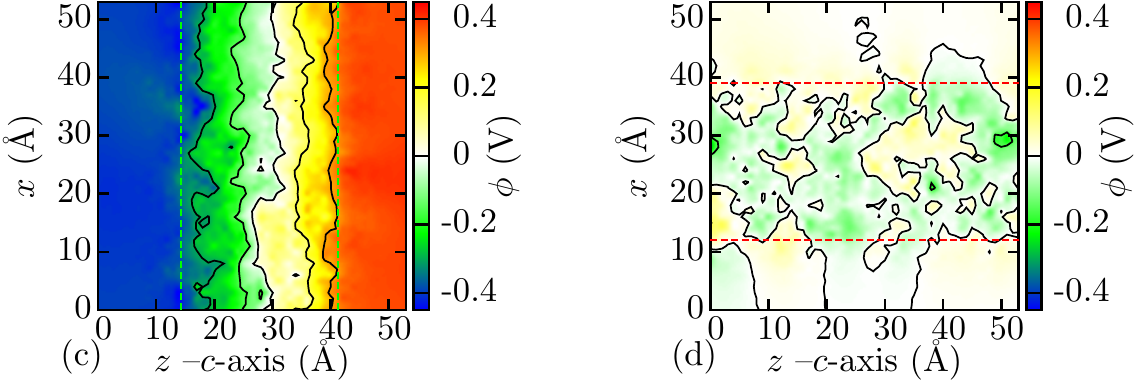}
\caption{(Color online) Sections in a plane parallel to the $c$-axis
of In$_{0.3}$Ga$_{0.7}$N/GaN QWs in polar and non-polar orientations.
The component of the polarization along the $c$-axis, $P_z$,
for the polar and non-polar structures is shown in (a) and (b),
respectively. The corresponding polarization potential is shown in
(c) and (d). The dashed lines indicate the approximate location of the
interfaces between well and barriers. The arrows in (a) and (b)
give the direction of the polarization in the $xz$ plane, as well as its
magnitude in the same
units as the arrow in the legend, which indicates 0.1 C/m$^2$. Solid
circles are Ga atoms and open circles are In atoms.}
\label{47}
\end{figure*}

Before applying the model to calculate the local polarization potential
in realistic structures, it is necessary to test its accuracy against
well established methods. An excellent test is the calculation of the
polarization potential in a capacitor-like structure. In such an example,
a layer of dielectric material (1) of thickness $h$, in which the polarization
$\textbf{P} = P_0 \, \hat{\textbf{z}}$ (where $\hat{\textbf{z}}$ is
a unit vector along the $z$ axis) is constant and perpendicular to
the neighboring interfaces,
is surrounded by two infinite layers of a dielectric material (2),
with a different dielectric constant, in which the polarization is zero.
An exact analytical solution to Poisson's equation can be obtained
for the latter case. If we assume the first interface is located at $z=0$,
the potential is given by
\begin{align}
\phi \left( \textbf{r} \right) = 
\frac{P_0}{2 \varepsilon_0 \varepsilon_r^{(1)}} \left( |z| - |z-h| \right),
\label{42}
\end{align}
where $\varepsilon_r^{(1)}$ is the dielectric constant of material (1).
Figure~\ref{43}(a) shows the potential profile as calculated exactly and
analytically using Eq.~(\ref{42}) for the special case in which
$P_0 = 0.1 \text{ C/m}^2$, $\varepsilon_r^{(1)} = 8.4$ and
$h = 30 \text{ \AA}$, which would be
typically the situation in an InGaN QW surrounded by GaN barriers
in which, for simplicity, the polarization has been switched off
in the barriers. Within this simplified continuum picture,
a spatial discretization of the current problem in a cubic grid of steps
$\Delta \approx 2 \text{ \AA}$, as discussed in the previous section,
creates an ensemble of point dipoles which are of similar size to the
ones encountered in typical InGaN QW situations.
The application of our dipole method to first order reflections
(see Appendix~\ref{241})
leads to a potential profile as in Fig.~\ref{43}(b). In that
figure, it can be observed how the potential changes brusquely in the
surroundings of the dipoles (the plane of the figure has been deliberately
chosen to be one that contains dipoles in it to dramatize this effect).
This is due to the fact that Eq.~(\ref{39})
is a valid solution for a distribution of charge
only if the position where the potential is calculated
is sufficiently far away from the location of the point dipole that
represents that distribution. We acknowledged this limitation in our
previous work and proposed a cutoff radius around $\textbf{r}$ for which
only the dipoles that obey the condition
$|\textbf{r} - \textbf{r}'| > r_\text{cutoff}$ are taken
into account.~\cite{caro_2012} The potential profile for the present
example and $r_\text{cutoff} = 1\text{ \AA}$ is shown in
Fig.~\ref{43}(c). Although this solution certainly improves the
results and leads to a much better agreement with the analytical solution,
it has the inconvenience of creating sharp transitions at the cutoff distances
around the dipoles. To complement this treatment, we have now substituted
the elimination of dipoles below the cutoff radius by a
Gaussian smearing of dipoles that obey the condition
$|\textbf{r} - \textbf{r}'| < r_\text{smear}$, as detailed in
Section~\ref{219} of Appendix~\ref{241}.
This solution leads to smoother potentials and
a much better agreement with the analytic solution for this test case, as
observed in Fig.~\ref{43}(d).

\section{Selected results for $\text{InGaN}$ quantum wells}\label{04}

Once the method for calculating the local polarization potential has
been established, we can turn our attention towards achieving a local
description of that quantity in relevant nanostructures. In the
present example, we look at InGaN/GaN QWs grown along polar and
non-polar\cite{*[{Non-polar QWs structures have precisely been
proposed as a possible solution to the built-in field issue in
nitride heterostructures, see }] [{.}] paskova_2008}
directions. Polar structures are grown along the $c$-axis, whereas in the
case of non-polar structures the $c$-axis lies within the growth
plane. In a macroscopic picture of the polarization,
there are no discontinuities in $\textbf{P}$
between the well and barriers in the non-polar case.
However, as we shall see, in a microscopic description
discontinuities occur locally, depending on local strain and composition.

Although we used in Section~\ref{07} DFT to optimize the atomic
positions of the supercells studied, such an approach is unaffordable
for large supercells, given both the computer time and memory usage required.
The usual approach to relax the atomic degrees of freedom in such cases
is to use a classical interatomic force method. For
tetrahedrally bonded compounds, Keating's valence force field (VFF)
model~\cite{keating_1966}
is by far the most popular.~\cite{pryor_1998,camacho_2010} Camacho and
Niquet have previously used a modified version of Keating's model,
adapted to the WZ crystal structure, to account for the deviation
of the $c/a$ ratio of lattice parameters with respect to its ideal
value.~\cite{camacho_2010} We have instead chosen an approach based
on Martin's VFF~\cite{martin_1972} that includes the electrostatic
interaction explicitly.~\cite{caro_2012} At a higher computational cost,
this model succeeds at predicting the deviation of the $c/a$ ratio
while maintaining the correct symmetry of the interatomic interactions. For
instance, the two-body interactions directed along the WZ $c$-axis
have the same functional form,
including the equilibrium bond length, as the other ones. This allows to
obtain a much more flexible set of potentials that are transferable between
similar polymorphs of the same compound, i.e. WZ and ZB
in this case. With our model we are able to predict elastic and
structural properties of binary and ternary nitrides in
excellent agreement with first-principles DFT calculations,
therefore providing solid grounds for using the supercells relaxed
using this method as high-quality input for the subsequent
local polarization calculation. An extensive
article with the details and validity of our method is currently
in preparation and will be published elsewhere.

Making use of the expressions derived throughout this chapter,
and the VFF just outlined, we have calculated the
local polarization for InGaN/GaN QWs with 30\% In content in both polar
and non-polar orientations, as shown in Figs.~\ref{47}(a) and (b),
respectively. Note that the component shown in the color code is
the component of the polarization along the $c$-axis. The corresponding
polarization potential is shown in Figs.~\ref{47}(c), for the polar
case, and (d), for the non-polar situation.
The polar structure shows a potential profile with the main features
of a capacitor-like structure, although significant fluctuations can also be
observed.
For a constant value of the polarization, i.e. with no local effects taken into
account, the isolines in Fig.~\ref{47}(c) would be perfectly parallel
to each other, as seen already in Fig.~\ref{43}. In the non-polar
case [Fig.~\ref{47}(d)] there are no main features in the potential
but only local effects.

Note that the non-polar QW situation
is similar to a bulk calculation in the sense that there are no macroscopic
polarization discontinuities, and the polarization potential landscape is
only affected by local effects. The importance of these local
effects will be highlighted in the next
section where we present tight-binding calculations of the electronic
structure of bulk InGaN alloys.

\section{Tight-binding model for electronic structure calculation}\label{05}

In this section we outline the ingredients for our electronic
structure calculations. We begin in Section~\ref{48} by
introducing the tight-binding (TB) model used to study the
band gap bowing in InGaN alloys. We first introduce
the TB model employed to describe the binary bulk materials InN and GaN.
We then outline how strain and built-in potential are
included in the description as well as how the TB model is
implemented to describe the ternary material InGaN.

\subsection{Binary bulk systems}\label{48}

To investigate the band gap bowing of ternary materials a
microscopic description of the system is required. An ideal solution
to this problem would be to perform DFT-based calculations.
However, standard DFT approaches fail to
provide an accurate description of the band gaps, especially for
systems with a small band gap.~\cite{moses_2011} As we have seen, standard
calculations within LDA or GGA
tend to predict a metallic phase for InN, while
experiments show a band gap of 0.6--0.7 eV.~\cite{wu_2009}
As we have previously discussed, HSE hybrid
functional DFT calculations~\cite{heyd_2003,heyd_2004} have attracted
considerable attention since within this framework one reduces
these band gap problems.~\cite{moses_2011} Even though standard
HSE-DFT calculations circumvent problems with the band gap in
general, these methods still underestimate the band gaps of InN, GaN
and AlN.~\cite{moses_2011} Especially, if one aims for a comparison
with experimentally determined transition energies and band gap
bowing parameters, an accurate description of the band gaps of the
binary compounds becomes important. Therefore, an approach is required
which reproduces effective masses, energetic positions of the
different valence bands (VBs) and conduction bands (CBs) and
additionally gives band gaps of the binary compounds in agreement
with experiment. On the other hand, this approach must also allow for
a microscopic description of the alloys. Such a description can be
achieved by pseudo-potential~\cite{chan_2010} or TB
calculations.~\cite{oreilly_2009} In the following we apply the TB
method to analyze the band gap bowing in wurtzite InGaN
alloys.

More specifically, we use a microscopic $sp^3$ TB
model. In this TB model the relevant electronic structure of anions
and cations is described by the
outermost valence orbitals, $s$, $p_x$, $p_y$ and $p_z$, and the
overlap of these basis orbitals is restricted to nearest neighbors.
Being only of the order of a few meV, we neglect the spin-orbit (SO)
coupling in the model. The inclusion of the SO coupling is straight
forward and detailed for example in Ref.~\onlinecite{schulz_2008}.
However, the crystal field (CF) splitting $\Delta_\text{cf}$
must be included in the model
since it is of significant importance for the accurate description
of the VB structure of III-N compounds. Values of $\Delta_\text{cf}$
lie in the range of 19--24 meV and 9--38 meV for InN
and GaN, respectively, while for AlN
$\Delta_\text{cf}=-230$.~\cite{yan_2010}

To include the CF splitting in our TB model we proceed in the
following way. As discussed in Ref.~\onlinecite{kobayashi_1983}, the small
CF splitting $\Delta_{\text{cf}}$ in a WZ crystal
differentiates the $p_z$ orbital from the $p_x$ and $p_y$ orbitals.
LDA pseudopotential calculations
suggest that for the studied materials the bulk CF splitting
should be modeled when using the TB method by taking a specific  
third-nearest-neighbor interactions into
account.~\cite{murayama_1994} The TB model we are using here considers only
nearest-neighbor hopping matrix elements and treats the four nearest-neighbor
atoms as equivalent. To account for the CF splitting within
the empirical $sp^3$ TB model with nearest-neighbor coupling, we
introduce the additional parameter $E(p_z,a)$ on the anion sites for
the on-site matrix elements of the $p_z$ orbitals. This additional
term is used to reproduce the splitting of the valence bands at the zone
center ($\Gamma$ point). Such an approach has also been applied for
CdSe QDs with a wurtzite structure.~\cite{leung_1998}
With four atoms per unit cell, the resulting Hamiltonian is a
$16\times16$ matrix for each $\mathbf{k}$ point. This Hamiltonian
parametrically depends on the different TB matrix elements, as for
example shown in Ref.~\onlinecite{kobayashi_1983}.

In general, the TB matrix elements are treated as parameters and are
determined by fitting the bulk TB band structure to DFT band
structures. In doing so, the TB parameters are designed to reproduce
the characteristic features of the DFT band structures, such as
energy gaps and splittings between different VBs and CBs. Here we
have performed HSE-DFT band structure calculations for InN and GaN
according to the guidelines given in Ref.~\onlinecite{yan_2009}.
We used a $\Gamma$-centered $6\times6\times4$
\textit{k} mesh, and cutoff energy of 600 eV for plane waves.
These are the same settings as have been used in Section~\ref{02} for the
calculation of the polarization-related parameters for the III-N compounds.
Recently, we
have used the same settings to perform HSE-DFT based calculations
for elastic constants in wurtzite InN, AlN and GaN.~\cite{caro_2012c}
These calculations gave elastic constants in very good agreement
with available experimental data.~\cite{caro_2012c,caro_2012d}
\begin{figure}[t]
\includegraphics{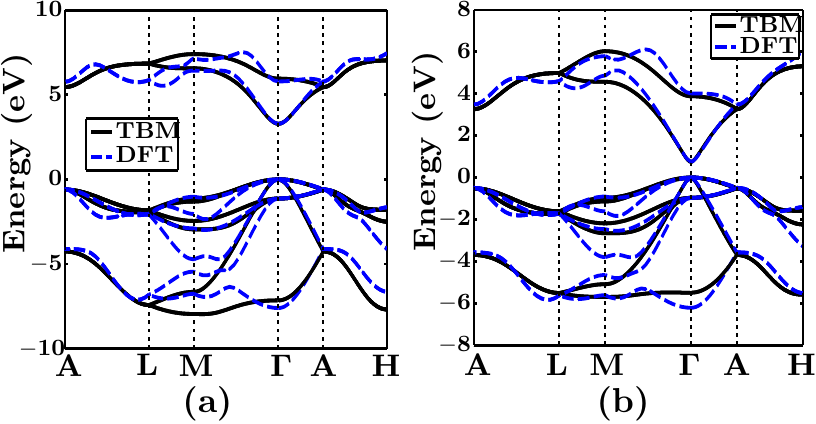}
\caption{(Color online) Bulk band structure of wurtzite (a) GaN  and
(b) InN obtained using HSE-DFT (dashed line)  and our $sp^3$ TB
model (solid line).}
\label{49}
\end{figure}
The HSE-DFT band structure serves as the reference for the TB
fitting procedure, for which we use a least-square
fitting at the $\Gamma$ point and $\mathbf{k}$ points
along the $\Gamma-A$ and $\Gamma-M$ directions, following the
guidelines given in
Ref.~\onlinecite{vogl_1983}. This ensures that the energetic positions
near the CB and VB edges as well as the curvature of the different TB
bands in the vicinity of the $\Gamma$ point are in good agreement
with the HSE-DFT calculations. Furthermore, using the guidelines of
Ref.~\onlinecite{vogl_1983}, chemical trends are also taken into account. The
resulting TB band structures in comparison with the HSE-DFT band
structure for InN and GaN are shown in Fig.~\ref{49}.

However, as discussed above, and for example in more detail in
Ref.~\onlinecite{moses_2011}, even HSE-DFT calculations
underestimate the bulk band gap. Since this quantity is of central
importance for a detailed comparison with experimental data, we
adjust our TB model to reproduce the experimental band gap. In order to do
so, here we
shift the on-site cation $s$-orbital energies.
This procedure affects mainly the CB edge
and bands energetically further away from the VB and CB edges. These
bands are of secondary importance for the description of the band gap
bowing in InGaN alloys. Table~\ref{50} summarizes the
resulting TB parameters.

\begin{table}[t]
\caption{Tight-binding parameters (in eV) for the nearest neighbor
$s p^3$ model of wurtzite InN and GaN. The notation of
Ref.~\onlinecite{kobayashi_1983} is used.}
\begin{ruledtabular}
\begin{tabular}{l c c}
& InN & GaN
\\
\hline
$E(s,a)$  & -11.92 & -10.62
\\
$E(p,a)$  & 0.49 & 0.82
\\
$E(p_z,a)$  & 0.46 & 0.79
\\
$E(s,c)$  & 0.48 & 0.91
\\
$E(p,c)$  & 6.53 & 6.68
\\
$V(s,s)$  & -1.61 & -5.97
\\
$V(x,x)$  & 1.79 & 2.34
\\
$V(x,y)$  & 4.83 & 5.47
\\
$V(sa,pc)$  & 1.89 & 4.09
\\
$V(pa,sc)$  & 6.14 & 8.67
\end{tabular}
\end{ruledtabular}
\label{50}
\end{table}

\subsection{Tight-binding description for alloys}\label{51}

In the framework of a TB model, the InGaN alloy is modeled on an
atomistic level. The TB parameters at each atomic site $\mathbf{R}$ of
the underlying wurtzite lattice are first set according to the bulk values
of the respective occupying atoms. While for the cation sites (Ga,
In) the nearest neighbors are always nitrogen atoms and there is no
ambiguity in assigning the TB on-site and nearest neighbor matrix
elements, this classification is more difficult for the nitrogen
atoms. In this case the nearest-neighbor environment is a
combination of In and Ga atoms. Here, we apply the widely used
approach of using weighted averages for the on-site energies
according to the number of neighboring In and Ga
atoms.~\cite{oreilly_2002,li_1992,boykin_2007} The hopping matrix elements
are chosen according to the values for InN or GaN.

In setting up the Hamiltonian, one must also include the local strain
$\epsilon_{ij}(\mathbf{r})$ and the total built-in potential $\phi$
to ensure an accurate description of the electronic properties of
the InGaN alloy. Several authors have shown that this can be done by
introducing on-site corrections to the TB matrix elements
$H_{l\mathbf{R}',m\mathbf{R}}$,~\cite{jancu_1998,boykin_2002} where
$\mathbf{R}$ and $\mathbf{R}'$ denote lattice sites and $l$ and $m$
are the orbital types. Therefore, we proceed in the following
way. The strain dependence of the TB matrix elements is included via
the Pikus-Bir Hamiltonian~\cite{winkelnkemper_2006,schulz_2010_b} as a
site-diagonal correction:
\begin{align}
H^\text{str}_{l\mathbf{R},m\mathbf{R}}=
\left(
\begin{array}{c c c c}
S_s & 0 & 0 & 0
\\
0 & S_x & S_{xy} & S_{xz}
\\
0 & S_{xy} & S_y & S_{yz}
\\
0 & S_{xz} & S_{yz} & S_z
\end{array}
\right) ,
\label{52}
\end{align}
with
\begin{align}
& S_s = a_{ct}(\epsilon_{11}+\epsilon_{22})+a_{cp}\epsilon_{zz} ,
\nonumber \\
& S_x = (D_2+D_4)(\epsilon_{11}+\epsilon_{22})+D_5(\epsilon_{11}-\epsilon_{22})
\nonumber \\
& \qquad +(D_1+D_3)\epsilon_{33} ,
\nonumber \\
& S_y = (D_2+D_4)(\epsilon_{11}+\epsilon_{22})-D_5(\epsilon_{11}-\epsilon_{22})
\nonumber \\
& \qquad +(D_1+D_3)\epsilon_{33},
\nonumber \\
& S_z = D_2(\epsilon_{11}+\epsilon_{11}) ,
\nonumber \\
& S_{xy} = 2 D_5 \epsilon_{12} ,
\nonumber \\
& S_{xz} = \sqrt{2} D_6 \epsilon_{13} ,
\nonumber \\
& S_{yz} = \sqrt{2} D_6 \epsilon_{23} ,
\end{align}
where the $D_i$ denote the VB deformation potentials, while
$a_{cp}$ and $a_{ct}$ are the CB deformation
potentials.~\footnote{Note that the quantities $a_2$ and $a_1$ given
by Vurgaftman and Meyer in their 2003 review article~\cite{vurgaftman_2003}
\textit{are not} the CB deformation potentials $a_{cp}$ and $a_{ct}$,
respectively. The quantities denoted by $a_1$ and $a_2$ in
Ref.~\onlinecite{vurgaftman_2003} are the \textit{band gap}
deformation potentials, e.g. $a_1=a_{cp}-D_1$ and $a_2=a_{ct}-D_2$.}
With this approach, the
relevant deformation potentials for the highest VB and lowest CB
states are included directly without any fitting procedure. In the
work described below, the deformation potentials for InN and GaN are
taken from HSE-DFT calculations.~\cite{yan_2009} Again, on the same
footing as in the case of the on-site energies for the nitrogen
atoms, we use weighted averages to obtain the strain-dependent
on-site corrections for In$_{x}$Ga$_{1-x}$N. Our
approach is similar to that used for the strain dependence in an
8-band \textbf{k}$\cdot$\textbf{p} model,~\cite{winkelnkemper_2006} but has
the benefit that the TB Hamiltonian still takes the correct
symmetry of the system into account, and is sensitive to In, Ga and
N atoms.

To obtain the local strain tensor $\epsilon_{ij}(\mathbf{r})$ at
each lattice site, we perform in a first step a relaxation of
the atomic positions in $\text{In}_x\text{Ga}_{1-x}$N supercells
based on the VFF outlined in Section~\ref{04}. From the relaxed atomic
positions,
we calculate $\epsilon_{ij}(\mathbf{r})$ according to the method in
Ref.~\onlinecite{pryor_1998} via:~\footnote{Note that the definition
of the strain matrix done by Pryor \textit{et al}.~\cite{pryor_1998}
is related to ours via a transposition operation.}
\begin{align}
\left(
\begin{array}{c c c}
\epsilon_{xx} & \epsilon_{xy} & \epsilon_{xz}
\\
\epsilon_{yx} & \epsilon_{yy} & \epsilon_{yz}
\\
\epsilon_{zx} & \epsilon_{zy} & \epsilon_{zz}
\end{array}
\right) = & \left(
\begin{array}{c c c}
R^0_{12,x} & R^0_{23,x} & R^0_{34,x}
\\
R^0_{12,y} & R^0_{23,y} & R^0_{34,y}
\\
R^0_{12,z} & R^0_{23,z} & R^0_{34,z}
\end{array}
\right)^{-1}
\nonumber \\
& \times \left(
\begin{array}{c c c}
R_{12,x} & R_{23,x} & R_{34,x}
\\
R_{12,y} & R_{23,y} & R_{34,y}
\\
R_{12,z} & R_{23,z} & R_{34,z}
\end{array}
\right) - \mathbbm{1},
\end{align}
where $\textbf{R}_{12}$, $\textbf{R}_{23}$ and $\textbf{R}_{34}$ are
the distorted tetrahedron edges, while $\textbf{R}^0_{12}$,
$\textbf{R}^0_{23}$ and $\textbf{R}^0_{34}$ are the ideal
tetrahedron edges. $\mathbbm{1}$ is the $3 \times 3$ identity matrix.
The built-in potential $\phi$ is likewise included as a site-diagonal
contribution in the TB Hamiltonian. This is also a widely used
approach.~\cite{ranjan_2003,saito_2002,schuh_2012}

\section{Results}\label{62}

In the following we use our TB model, including local strain and
built-in potentials to analyze the band gap bowing of InGaN. We
outline the procedure for TB supercell calculations in
Section~\ref{53}, while in Section~\ref{54} we
compare our theoretical results for transition energies and band gap
bowing parameters against experimental and other theoretical data.
The impact of local alloy composition, local strain and local
built-in potential on the CB and VB edges of InGaN alloys is discussed in
Section~\ref{55}.

\subsection{TB supercell calculations for InGaN}\label{53}

In the following, all calculations are performed on supercells
containing approximately 12,000 atoms, with periodic boundary
conditions applied. A large number of atoms are included in the supercell
to suppress the influence of finite-size supercell
effects. We assume that InGaN is a random alloy, following recent
experimental indication.~\cite{galtrey_2007,humphreys_2007}
For each In concentration we have
performed calculations with five different microscopic
configurations, where the In atoms are placed randomly in the
supercell. We calculate the band gap $E_g(x)$ as a configurational
average, i.e.
\begin{align}
E_g(x) = \frac{1}{N} \sum\limits_{i=1}^{N} \left[ E^{i}_\text{CB}(x)
- E^{i}_\text{VB}(x) \right],
\end{align}
where $i$ denotes the microscopic configuration and
$E^{i}_\text{CB}$ ($E^{i}_\text{VB}$) is the corresponding CB (VB)
edge. The number of configurations is given by $N$.

\subsection{Band gap bowing in InGaN: Comparison with experiment}\label{54}

\begin{figure}
\includegraphics{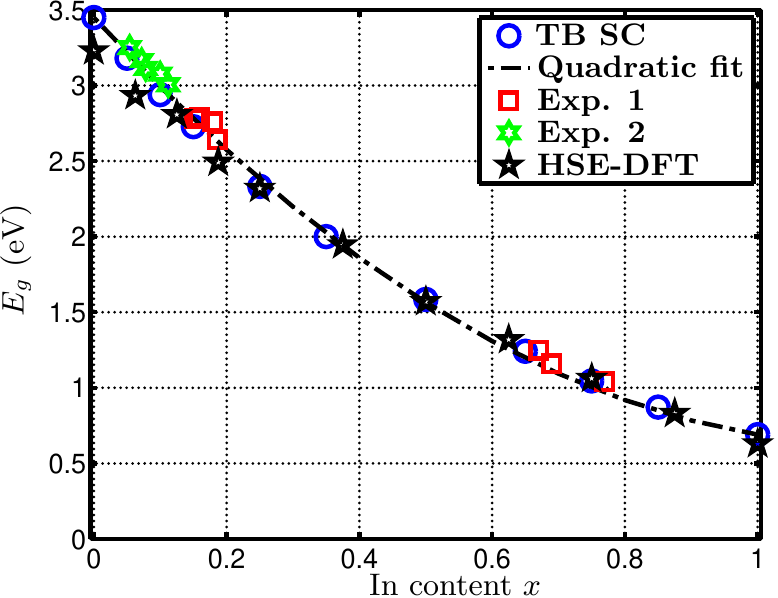}
\caption{(Color online) Band gap $E_g$ of
In$_x$Ga$_{1-x}$N as a function of the In content $x$.
Our TB supercell calculations (TB SC) are compared with experimental and
theoretical data. The dashed dotted line indicates the quadratic
fit, Eq.~(\ref{57}), to the TB data. Experimental data are taken from
Sakalauskas \textit{et al}.~\cite{sakalauskas_2012} and Schley \textit{et
al}.~\cite{schley_2007} (Exp. 1) and from McCluskey \textit{et
al}.~\cite{mccluskey_2003} (Exp. 2). Theoretical HSE-DFT data taken
from Moses \textit{et al}.~\cite{moses_2011}}
\label{56}
\end{figure}

Figure~\ref{56} shows TB supercell calculation results
(open blue circles) for the band gap $E_g$ of
In$_{x}$Ga$_{1-x}$N as a function of the In content $x$.
The TB results are compared to recent experimental
data~\cite{mccluskey_2003,schley_2007,sakalauskas_2012} and HSE-DFT
calculations~\cite{moses_2011}. From Fig.~\ref{56} one can
infer that our TB results are in excellent agreement with the
HSE-DFT results for In contents above 15\%--20\% ($x>0.15-0.2$). For values
below 15\% the fact that the HSE-DFT calculations underestimate the
band gap of GaN becomes important. However, in this composition
regime ($x<0.2$), our TB results are in very good agreement with the
experimental data, cf. Fig.~\ref{56}. Also in the high In
content regime ($x>0.5$) the TB data is in very good agreement with the
experimental data. text: We have applied this model to AlInN too,
also showing an excellent agreement with recent experimental
data.~\cite{schulz_2013}

For the design of In$_x$Ga$_{1-x}$N based optoelectronic
devices, knowledge about the behavior of the band gap $E_g$ with
composition $x$ is of central importance. Usually the dependence of
$E_g$ on $x$ is described by a quadratic function in $x$, involving
the energy gaps of InN ($E_g^\text{InN}$), GaN ($E_g^\text{GaN}$)
and a bowing parameter $b$:
\begin{align}
E_g = x E_g^\text{InN} + (1-x) E_g^\text{GaN} - b (1-x) x.
\label{57}
\end{align}
Commonly, the band gap bowing parameter $b$ of InGaN is assumed to be
composition \textit{independent}.~\cite{vurgaftman_2003} We start with this
assumption and denote the composition independent bowing parameter
by $\tilde{b}$. In doing so we find a bowing parameter
$\tilde{b} \approx 2$~eV. Experimentally determined bowing parameters
scatter quite significantly, ranging from 1.43 eV to 2.8 eV.
Theoretical values for $\tilde{b}$ range from 1.36 to 5.14 eV.
Compared to both theory and experiment our reported value of
$\tilde{b} \approx 2$~eV is therefore within the range of the
reported literature values.
However, it has been suggested~\cite{wetzel_1998,mccluskey_1998}
that the bowing
parameter of $\text{In}_{x}\text{Ga}_{1-x}$N alloys is composition
dependent. Based on HSE-DFT calculations for special quasirandom
structures (SQSs), Moses \textit{et al}.~\cite{moses_2011} found that
$b$ ranges in In$_x$Ga$_{1-x}$N from 2.29~eV
($x = 0.0625$) to 1.14~eV ($x = 0.875$). Gorczyca \textit{et
al}.~\cite{gorczyca_2009} used LDA+C calculations to analyze $b$ in
In$_x$Ga$_{1-x}$N alloys. The authors considered two
types of alloys, i.e. i) alloys with uniformly distributed In atoms
in a 32-atom supercell and ii) alloys with all In atoms clustered.
In case i) Gorczyca \textit{et al}.~\cite{gorczyca_2009} reported that $b$
ranges from 1.7~eV (large $x$) to 2.8~eV (small $x$). For $x = 0.5$
the authors found for the uniform case $ b(0.5) = 2.1$~eV. Looking at
case ii), the clustered alloy, band gap bowing values between
2.5~eV (large $x$) and 6.5~eV have been reported, with $ b(0.5) =3 .9$~eV.
Based on our random TB supercell calculations, we find that our
bowing parameter shows a strong composition dependence. The TB
results for $b$ are summarized in Table~\ref{58}.
Here, the values for $b$ range from 1.78~eV (large $x$) to 2.77~eV
(small $x$). At $x = 0.5$ we find $b = 1.94$~eV. Therefore, our results
are close to the results obtained from the LDA+C calculations in the
case of an uniform alloy (see above).

To shed more light on the composition dependence of $b$, we investigate
in a second step how the CB and VB edge behave as a function of the In
content $x$. These quantities are also of great interest for the
design of InGaN/GaN based optoelectronic devices, since the CB and VB
edge energies in InGaN affect the confinement energies of electron and hole
wave functions. Here, to calculate the bowing parameters $b^\text{CB}(x)$ and
$b^\text{VB}(x)$ for the CB and VB edge, respectively, we use:
\begin{align}
E_\text{CB} = & \left(E^\text{InN}_g + \Delta E_\text{VB} \right) x
+ E^\text{GaN}_g( 1-x) - b^\text{CB} (1-x) x
\nonumber \\
E_\text{VB} = & \Delta E_\text{VB}x - b^\text{CB} (1-x) x ,
\label{59}
\end{align}
where $E_\text{CB}$ and $E_\text{VB}$ are the CB and VB edges,
respectively. These quantities are obtained from our TB SC
calculations. The VB offset is denoted by $\Delta E_\text{VB}$ and
taken from HSE-DFT data in Ref.~\onlinecite{moses_2011}. Here,
$b^\text{CB}(x)$ and $b^\text{VB}(x)$ are composition-dependent
fitting parameters to reproduce $E_\text{CB}$ and $E_\text{VB}$,
respectively. The resulting composition-dependent values for
$b^\text{CB}(x)$ and $b^\text{VB}(x)$ are summarized in
Table~\ref{58}. From this table one infers that, while
$b^\text{VB}$ is almost composition independent, $b^\text{CB}$
varies significantly with $x$. Consequently, the composition
dependence of the band gap bowing $b$ arises mainly from the
composition dependence of the CB edge. This result is in agreement
with the HSE-DFT findings of Ref.~\onlinecite{moses_2011}. Therefore,
when modeling InGaN based heterostructures in the framework of a
continuum description, such as \textbf{k}$\cdot$\textbf{p} theory,
the composition dependent bowing of the band edges should be taken
into account in order to
achieve a realistic description of these systems.

To extend the analysis of the band edges in InGaN alloys further we
focus in the next section on the impact of local composition, local
strain and local built-in potentials on the CB and VB edge,
respectively.

\begin{table}[t]
\squeezetable
\caption{Band gap bowing parameter $b(x)$ of
In$_x$Ga$_{1-x}$N as a function of the In content $x$.}
\begin{ruledtabular}
\begin{tabular}{l c c c c c c c c c}
$x$ & 5\% & 10\% & 15\% & 25\% & 35\% & 50\% & 65\% & 75\% & 85\%
\\
\hline 
$b$ (eV) & 2.77 & 2.6 & 2.42 & 2.28 & 2.13 & 1.94 &
1.82 & 1.78 & 1.82 
\\
$b^\text{CB}$ (eV) & 1.74 & 1.56 & 1.43 & 1.26 & 1.13 & 0.92 & 0.85
& 0.81 & 0.78
\\
$b^\text{VB}$ (eV) & -1.03 & -1.04 & -0.99 & -1.02 & -1.00 & -1.02 & -0.97
& -0.97 & -1.04
\end{tabular}
\end{ruledtabular}
\label{58}
\end{table}

\subsection{Impact of local composition, strain and built-in
potential on CB and VB edges in InGaN}\label{55}

\begin{figure}
\includegraphics{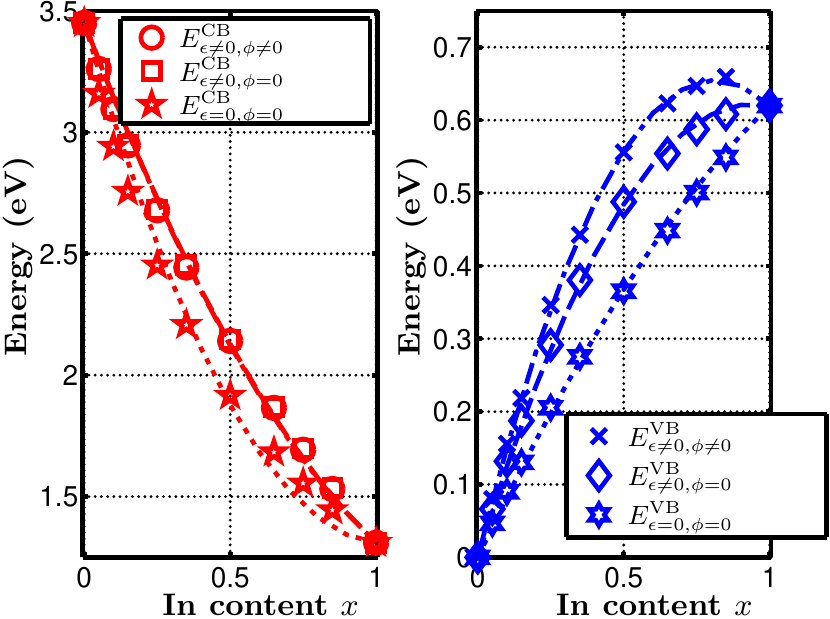}
\caption{(Color online) CB (left) and VB (right) edges of
In$_x$Ga$_{1-x}$N as a function of the In content $x$.
Dotted line: Fit to data without strain and built-in potential
($\text{VB/CB}_{\epsilon=0,\phi=0}$); Dashed line: Fit to data with
strain but without built-in potential
($\text{VB/CB}_{\epsilon\neq0,\phi=0}$); Dashed-dotted line: Fit to
data with strain and built-in potential
($\text{VB/CB}_{\epsilon\neq0,\phi\neq0}$).}
\label{60}
\end{figure}

In the previous section we have discussed the composition dependence
of the bowing parameters $b^\text{CB}(x)$ and $b^\text{VB}(x)$ of
the CB and VB edges, respectively. These calculations included local
strain and built-in potential effects due to alloy fluctuations.
Here we analyze in more detail how the different contributions from
pure alloy fluctuations, local strain and built-in potential effects
influence the CB and VB edge energies. Figure~\ref{60} shows the
CB (left) and VB (right) edge energies as a function of the In content $x$.

In a first step, to study the impact of the alloy fluctuations only,
we neglect the local strain and built-in field effects in the TB SC
calculations of the band edges
($E^\text{CB}_{\epsilon\neq0,\phi\neq0}$,
$E^\text{VB}_{\epsilon\neq0,\phi\neq0}$). As
Fig.~\ref{60} shows, in the absence of strain and
built-in potential, $E^\text{VB}_{\epsilon=0,\phi=0}$ varies almost
linearly with the In content $x$, while
$E^\text{CB}_{\epsilon=0,\phi=0}$ shows a strong non-linear
behavior. Using Eqs.~(\ref{59}), we can determine
$\tilde{b}^\text{CB}$ and $\tilde{b}^\text{VB}$
for the case of $\epsilon=0, \phi=0$.
The results for the composition-independent
bowing parameters $\tilde{b}^\text{CB}$
and $\tilde{b}^\text{VB}$
are summarized in Table~\ref{61}. When including local
strain effects but neglecting the local built-in potential,
$E^\text{CB}_{\epsilon\neq0,\phi=0}$ is shifted to higher energies
over the whole composition range due to hydrostatic strain in the system, (cf.
Fig.~\ref{60}). This reduces the CB edge bowing
parameter $\tilde{b}^\text{CB}$ by a factor of two compared to the situation
without strain and built-in potential effects (cf.
Table~\ref{61}). When looking at the behavior of
$E^\text{VB}_{\epsilon\neq0,\phi=0}$ in comparison to
$E^\text{VB}_{\epsilon=0,\phi=0}$ we find also a shift to higher
energies resulting from biaxial compressive strain.
However, in this case the magnitude of the VB edge bowing
parameter $\tilde{b}^\text{VB}$ is increased by a factor of three compared
to the situation without strain and built-in potential (cf.
Table~\ref{61}). When including both local strain and
built-in potential effects, $E^\text{CB}_{\epsilon\neq0,\phi\neq0}$
in comparison to $E^\text{CB}_{\epsilon\neq0,\phi=0}$ is almost
unaffected. This is also reflected in the data for the CB edge
bowing parameter $\tilde{b}^\text{CB}$ shown in Table~\ref{61}.
For the VB edge this is not the case. Here, the local built-in potential
significantly modifies the VB edge, as seen in
Fig.~\ref{60}. Moreover, due to local built-in
potential effects, $E^\text{VB}_{\epsilon\neq0,\phi\neq0}$ exceeds
the InN/GaN VB offset $\Delta E_\text{VB}$. The consequence of this behavior
would be that In$_x$Ga$_{1-x}$N on InN would be a type-II heterostructure
for $x \gtrsim 0.6$.

\begin{table}[t]
\caption{Overall band gap ($\tilde{b}^\text{full}$), CB
($\tilde{b}^\text{CB}$) and
VB $\tilde{b}^\text{VB}$ bowing parameters. The results are shown in the
absence of strain and built-in potential ($\epsilon=0, \phi=0$), in
absence of the built-in potential but in the presence of strain
($\epsilon\neq0, \phi=0$) and finally with strain and built-in
potential included ($\epsilon\neq0, \phi\neq0$).}
\begin{ruledtabular}
\begin{tabular}{l c c c}
& $\tilde{b}$ (eV) & $\tilde{b}^\text{CB}$ (eV) & $\tilde{b}^\text{VB}$ (eV)
\\
\hline
$\epsilon=0, \phi=0$ & 2.24 & 2.01 & -0.23
\\
$\epsilon\neq0, \phi=0$ & 1.70 & 1.01 & -0.69
\\
$\epsilon\neq0, \phi\neq0$ & 2.02 & 1.03 & -0.99
\end{tabular}
\end{ruledtabular}
\label{61}
\end{table}

This difference in the behavior of the CB and VB edges can be
attributed in part to the differences in the effective masses. Compared to
the VB, the effective mass of the CB edge is
small.~\cite{rinke_2008,schulz_2010_b} Therefore, in the regime of large
$x$ (high In content), the randomly distributed In atoms can form
QD-like regions that lead to a localization of VB wave functions
since the local compressive strain favors this
behavior.~\cite{schulz_2012c} Therefore, we observe a strong increase
in the magnitude of $\tilde{b}^\text{VB}$ when including strain effects, cf.
Fig.~\ref{60} and Table~\ref{61}. In
contrast, the compressive hydrostatic strain in these regions leads
to a weaker localization of the CB wave functions and a shift to
higher energies,~\cite{schulz_2012c} as observed in
Fig.~\ref{60}. However, since the CB wave
functions are only weakly localized in the QD-like regions due to
strain effects and the low effective masses, the local built-in
potential is of secondary importance for the CB edge. However,
originating from the much stronger VB wave function localization,
as in a ``real'' nitride-based QD, the built-in potential further
increases the localization and leads to a pronounced shift to higher
energies.~\cite{williams_2009} As seen for example in experiments on
$c$-plane GaN/AlN QDs, due to the presence of the built-in potential
the measured photoluminescence (PL) energy drops below the GaN band
gap value.~\cite{simon_2003}

\section{Summary}\label{06}

We have presented a complete theory of local electric
polarization in the linear piezoelectric limit. The connection between
the local polarization and local internal strain is obtained in an elegant
manner through the use of Born effective charges and internal strain
parameters.
We have validated the theory for the highly ionic III-N wurtzite compounds,
demonstrating a high degree of agreement between our model and
Berry-phase calculations.
We have cast these local effects in the form of a local piezoelectric tensor,
which helps to highlight the importance
of local strain and tetrahedron orientation
on the polarization field and potential.
In addition to this, we have obtained a consistent
series of polarization-related \textit{ab initio} parameters for the group-III
nitrides.

We have also presented a point dipole method for the calculation of the local
polarization potential that overcomes resolution problems encountered
when solving directly Poisson's equation.
The method involves the discretization of the polarization field as
a series of point dipoles. The accuracy of the method has been tested
against a well known problem with analytical solution.
As an example,
we have applied our theory and methodology to study the local
polarization and local polarization potential in polar and non-polar
InGaN/GaN QW structures, where we have observed large local fluctuations
in both quantities.

Finally, we have presented a tight-binding model that allows us to
take into account local alloy effects, including local strain and
the local polarization potential discussed throughout the paper. With this
model we have calculated the composition dependence of the band gap
of InGaN and provided composition-dependent bowing parameters for the band gap
and both the conduction and valence band edge energies. Furthermore,
we have shown that the local polarization potential has a strong
influence on wave function localization effects in the valence band of this
material.

\begin{acknowledgments}
M.~A.~C. would like to thank Vincenzo Fiorentini and David Vanderbilt
for very useful discussions on the practicalities of Berry-phase
calculations.
S.~S. would like to thank Muhammad Usman for valuable discussions.
This work was carried out with the
financial support of Science Foundation Ireland under project number
10/IN.1/I2994.
S.~S. also acknowledges financial support from the European Union Seventh
Framework Programme (ALIGHT FP7-280587).
\end{acknowledgments}

\appendix

\section{Example of the calculation of a local piezoelectric coefficient}
\label{240}

To illustrate how the calculation of the local piezoelectric tensor
in terms of the internal strain parameters is done,
we give here the details of the calculation for
$e_{15}^{*,\text{A}}$.
The expression of $e_{ij}^{*,X}$ for $e_{15}^{*,\text{A}}$ is simplified to
\begin{align}
e_{15}^{*,\text{A}} = e_{15}^{(0)} - \frac{e \mathcal{Z}_1^\text{A}}{\sqrt{3}
{a_0}^2 c_0} \left(
\frac{\partial \mu_1^\text{A}}{\partial \epsilon_5} - \frac{1}{2}
\mu_{3,0}^\text{A} \right),
\label{239}
\end{align}
where we have made use of the Voigt relation $\partial \epsilon_{13} /
\partial \epsilon_5 = 1 / 2$.
$\mu_{3,0}^\text{A} = 4 (u_0 - 3/8 ) c_0$ is given by the WZ internal
parameter, and $\mathcal{Z}_1^\text{A} \equiv \mathcal{Z}_1$ for A
being a cation.
We need to calculate $\mu_1^\text{A}$.
Looking at Fig.~\ref{27}, it is clear that
the nearest neighbors of A are B, which we label 1, and three periodic
replicas of D contained in a plane below A, which we label 2--4.
If A is fixed at the origin, $\textbf{r}_\text{A} = (0,0,0)$, then
the distances of the different nearest neighbors from A are given by:
\begin{align}
& \boldsymbol{\ell}^1 = \textbf{r}_\text{B},
\nonumber \\
& \boldsymbol{\ell}^2 = \textbf{r}_\text{D} - \textbf{c},
\nonumber \\
& \boldsymbol{\ell}^3 = \textbf{r}_\text{D} - \textbf{c} - \textbf{a},
\nonumber \\
& \boldsymbol{\ell}^4 = \textbf{r}_\text{D} - \textbf{c} - \textbf{b},
\end{align}
where \textbf{a}, \textbf{b} and \textbf{c} are the (strained)
lattice vectors of the unit cell.
Since for this example we are interested in $e_{15}^{*,\text{A}}$ only,
we set all the strain components to zero except for $\epsilon_5 =
2\epsilon_{13}$. Following all the definitions given
in Ref.~\onlinecite{caro_2012c} (with exchanged notation
$\epsilon_{13} \leftrightarrow \epsilon_{xz}$), we can write:
\begin{align}
& \textbf{r}_\text{B} = \left[ u_0 c_0 \epsilon_{13}, 0, u_0 c_0 \right]
+ \left[ \zeta_1 c_0 \epsilon_{13}, 0, 0 \right],
\nonumber \\
& \textbf{r}_\text{D} = \Bigg[ \frac{a_0}{2} + \left( \frac{1}{2} +
u_0 \right) c_0 \epsilon_{13}, \frac{\sqrt{3} a_0}{6},
\left( \frac{1}{2} + u_0 \right) c_0
\nonumber \\
& \qquad \quad + \frac{a_0}{2} \epsilon_{13} \Bigg]
+ \left[ \zeta_1 c_0 \epsilon_{13}, 0, 0 \right],
\nonumber \\
& \textbf{a} = \left[ a_0, 0, a_0 \epsilon_{13} \right],
\qquad
\textbf{b} = \left[ \frac{a_0}{2}, \frac{\sqrt{3} a_0}{2}, \frac{a_0}{2}
\epsilon_{13} \right],
\nonumber \\
& \textbf{c} = \left[ c_0 \epsilon_{13}, 0, c_0 \right].
\end{align}
To obtain $\mu_1^\text{A}$ we sum over nearest-neighbor distances:
\begin{align}
\mu_1^\text{A} = & \sum\limits_{\alpha = 1}^{4} \ell_1^\alpha
= 4 u_0 c_0 \epsilon_{13} - \frac{3}{2} c_0 \epsilon_{13} + 4 \zeta_1 c_0
\epsilon_{13}.
\end{align}
The last term of \eq{239} therefore reduces to
\begin{align}
\frac{\partial \mu_1^\text{A}}{\partial \epsilon_5} - \frac{1}{2}
\mu_{3,0}^\text{A} = & 2 \left( u_0 - \frac{3}{8} \right) c_0 + 2 \zeta_1
c_0 - 2 \left( u_0 - \frac{3}{8} \right) c_0
\nonumber \\
= & 2 \zeta_1 c_0,
\end{align}
which leads to the final result:
\begin{align}
e_{15}^{*,\text{A}} = e_{15}^{(0)} - \frac{2 e \mathcal{Z}_1}{\sqrt{3}
{a_0}^2} \zeta_1.
\end{align}

\section{Point dipole method for the calculation of local polarization
potentials}
\label{241}

\begin{figure}[t]
\includegraphics{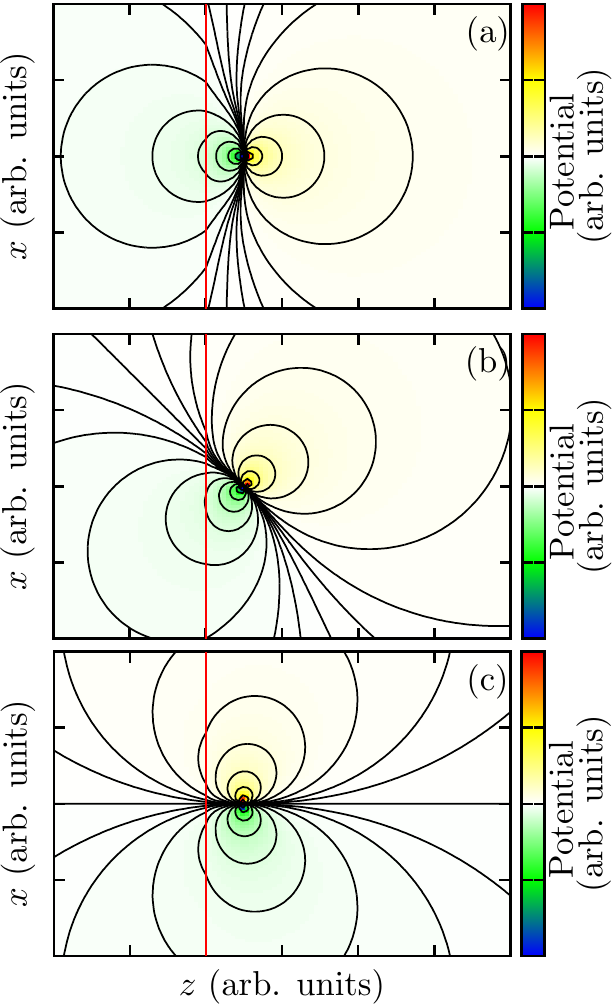}
\caption{(Color online)
Potential profiles for three dipole orientations in the case
of only one planar interface (indicated by the vertical line)
and two different dielectric constants. The potential isolines are chosen
so they decay exponentially.}
\label{41}
\end{figure}

Proceeding in a
similar manner to the one employed by Jackson for a point
charge,~\cite{jackson_1999} we can obtain the exact analytic solution
for the potential due to a point dipole when only one interface is
present, as schematically shown in Fig.~\ref{40}:
\begin{figure}[t]
\includegraphics{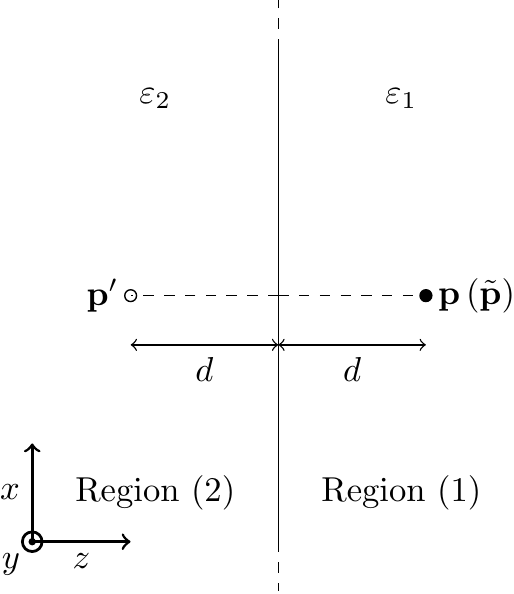}
\caption{Schematic representation
of the two media with different dielectric constant
and point dipole $\textbf{p}$ problem. The image
dipoles $\textbf{p}'$ and $\tilde{\textbf{p}}$ are needed in order
to solve it.}
\label{40}
\end{figure}
\begin{align}
&\phi_{\textbf{p}}^{(1)} \left( \textbf{r} \right) = \frac{1}{4 \pi \varepsilon_0
\varepsilon_1} \frac{\textbf{p} \cdot \left( \textbf{r} - \textbf{r}_\textbf{p}
\right)}{|\textbf{r} - \textbf{r}_\textbf{p}|^3} +
\frac{1}{4 \pi \varepsilon_0
\varepsilon_1} \frac{\textbf{p}' \cdot \left( \textbf{r} -
\textbf{r}_{\textbf{p}'}
\right)}{|\textbf{r} - \textbf{r}_{\textbf{p}'}|^3},
\nonumber \\
&\phi_{\textbf{p}}^{(2)} \left( \textbf{r} \right) = \frac{1}{4 \pi \varepsilon_0
\varepsilon_2} \frac{\tilde{\textbf{p}} \cdot \left( \textbf{r} -
\textbf{r}_{\tilde{\textbf{p}}}
\right)}{|\textbf{r} - \textbf{r}_{\tilde{\textbf{p}}}|^3},
\end{align}
with
\begin{align}
&\textbf{p}' = \frac{\varepsilon_1 - \varepsilon_2}
{\varepsilon_1 + \varepsilon_2} \left[ p_x, p_y, -p_z \right],
\qquad
\textbf{r}_{\textbf{p}'} = \left[ x_\textbf{p}, y_\textbf{p}, z_\textbf{p}
- 2d \right],
\nonumber \\
&\tilde{\textbf{p}} = \frac{2 \varepsilon_2}
{\varepsilon_1 + \varepsilon_2} \left[ p_x, p_y, p_z \right],
\qquad
\textbf{r}_{\tilde{\textbf{p}}} = \textbf{r}_{\textbf{p}},
\end{align}
where $\textbf{p}'$ is the image dipole used, together with the original dipole
$\textbf{p}$, for the calculation of the potential
$\phi_{\textbf{p}}^{(1)} \left( \textbf{r} \right)$ in region (1) and
$\tilde{\textbf{p}}$ is the image dipole used for the calculation of the
potential $\phi_{\textbf{p}}^{(2)} \left( \textbf{r} \right)$ in region
(2). Their positions are given by $\textbf{r}_{\textbf{p}'}$ and
$\textbf{r}_{\tilde{\textbf{p}}}$, respectively. The results for a test
dipole of arbitrary magnitude when one of the materials has a dielectric
constant twice as big as that of the material in which the dipole is contained
are shown in Fig.~\ref{41}(a{\textendash}c)
for three different orientations of the dipole.

The calculation of the potential when a second interface is included is
more complicated, as additional mirror images have to be added to balance
the two initial image dipoles about each interface. As a result, an
infinite number of reflections (and hence, image dipoles) have to be
considered in order to obtain the exact form of the potential. These
reflections up to third order are shown in Fig.~\ref{179}.
The treatment for a point charge in such a situation has been already done
by Barrera.~\cite{barrera_1978} For the case of a point dipole, we find the
expressions to be similar although the transformation of the point dipole is
somehow different compared to the point charge due to the vector nature
of the former. Details of our treatment and expressions for the three-media
case are given in the next section.

\subsection{Point dipole solution for the three-dielectric problem}
\label{185}

\begin{figure*}[t]
\includegraphics{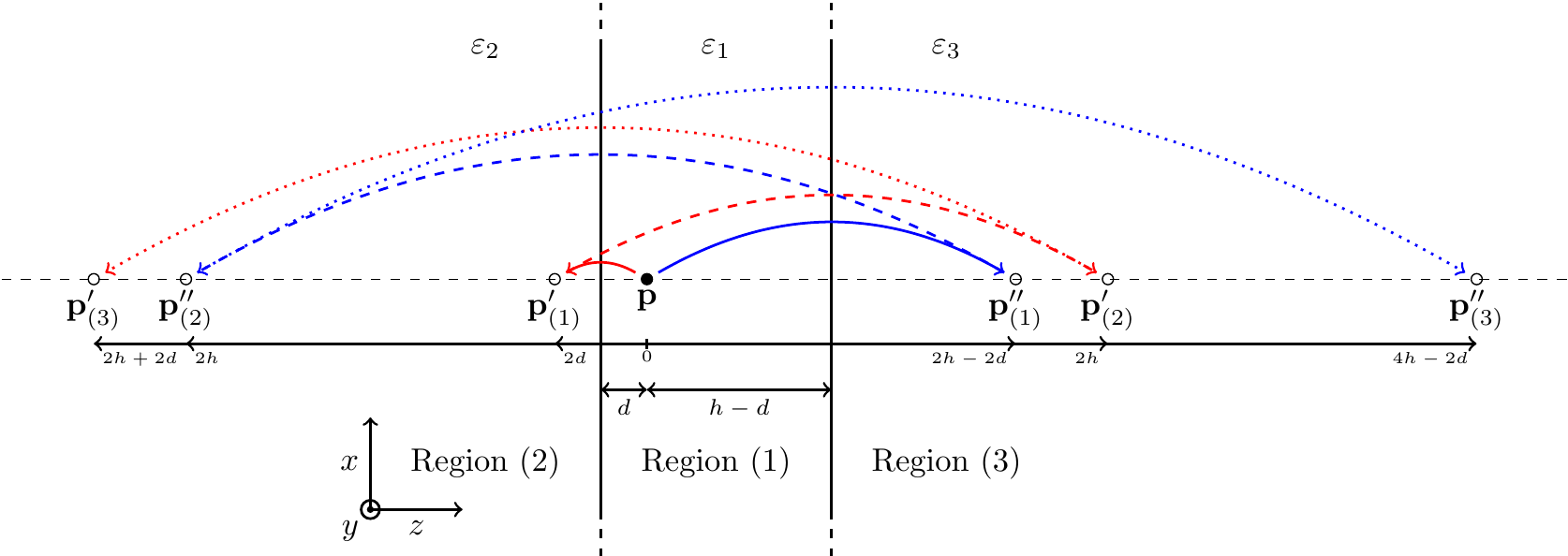}
\caption{Reflection of image dipoles up to third order in a
three-dielectric set up. Each of the
reflection sequences is denoted by a different color: the sequence starting
to the left and originating the series $\textbf{p}_{(n)}'$ is coloured in red
whereas the sequence starting to the right and originating the series
$\textbf{p}_{(n)}''$ is in blue. Solid lines indicate first order reflections,
dashed lines indicate second order reflections and dotted lines indicate
third order reflections.}
\label{179}
\end{figure*}

Building on the description made by Barrera for point
charges in a three-dielectric configuration,~\cite{barrera_1978}
we give here the analogous solution for point dipoles. The reflections
necessary to construct the image point dipoles are illustrated in
Fig.~\ref{179}.
Following the convention of Fig.~\ref{179}, where $d$ is the
distance from the dipole to the left side interface and $h$ is the distance
between the two interfaces, we can obtain a set of rules for the form
of the image charges $\textbf{p}_{(n)}'$ and $\textbf{p}_{(n)}''$, being
the $n\text{th}$ reflections of $\textbf{p}$ starting at left
and right, respectively.
These rules can be written as the following expressions. For the position
of the image dipoles:
\begin{align}
z_{\textbf{p}_{(2n-1)}'} &= z_{\textbf{p}} - \left[ n \, 2d + \left( n - 1
\right) \left( 2h - 2d \right) \right],
\nonumber \\
z_{\textbf{p}_{(2n)}'} &= z_{\textbf{p}} + \left[ n \, 2d + n
\left( 2h - 2d \right) \right],
\nonumber \\
z_{\textbf{p}_{(2n-1)}''} &= z_{\textbf{p}} + \left[ n \left( 2h - 2d \right)
+ \left( n - 1 \right) 2d \right],
\nonumber \\
z_{\textbf{p}_{(2n)}''} &= z_{\textbf{p}} - \left[ n \left( 2h - 2d \right)
+ n 2d \right],
\end{align}
and for the value of the image dipoles:
\begin{align}
\textbf{p}_{(2n-1)}' &= \left[ p_x, p_y, - p_z \right] \left(
\frac{\varepsilon_1 - \varepsilon_2}{\varepsilon_1 + \varepsilon_2}
\right)^n \left(
\frac{\varepsilon_1 - \varepsilon_3}{\varepsilon_1 + \varepsilon_3}
\right)^{n-1},
\nonumber \\
\textbf{p}_{(2n)}' &= \left[ p_x, p_y, p_z \right] \left(
\frac{\varepsilon_1 - \varepsilon_2}{\varepsilon_1 + \varepsilon_2}
\right)^n \left(
\frac{\varepsilon_1 - \varepsilon_3}{\varepsilon_1 + \varepsilon_3}
\right)^n,
\end{align}
for the first series and
\begin{align}
\textbf{p}_{(2n-1)}'' &= \left[ p_x, p_y, - p_z \right] \left(
\frac{\varepsilon_1 - \varepsilon_2}{\varepsilon_1 + \varepsilon_2}
\right)^{n-1} \left(
\frac{\varepsilon_1 - \varepsilon_3}{\varepsilon_1 + \varepsilon_3}
\right)^n,
\nonumber \\
\textbf{p}_{(2n)}'' &= \textbf{p}_{(2n)}',
\end{align}
for the second series, with $n \in \mathbb{N}$.
\begin{figure*}[t]
\includegraphics{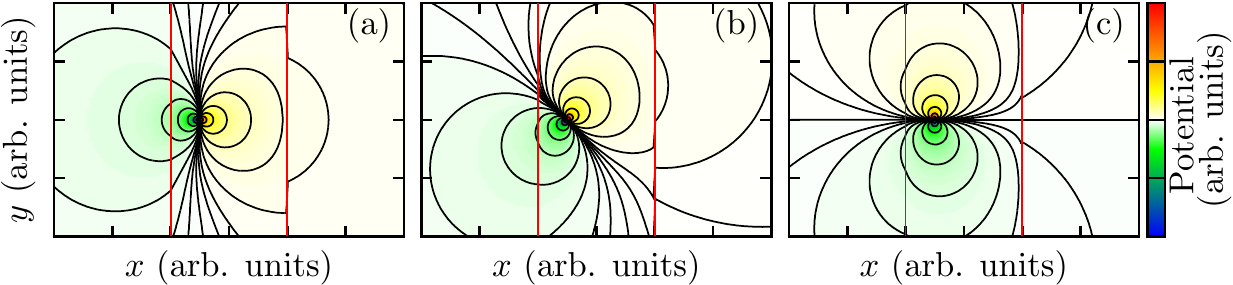}
\includegraphics{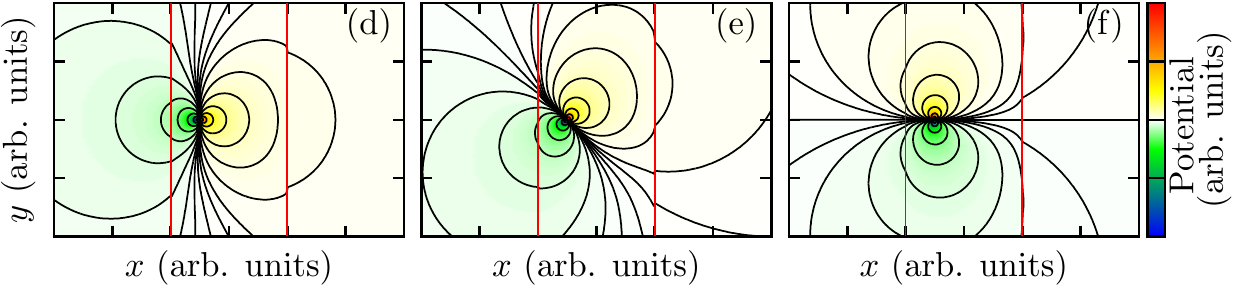}
\includegraphics{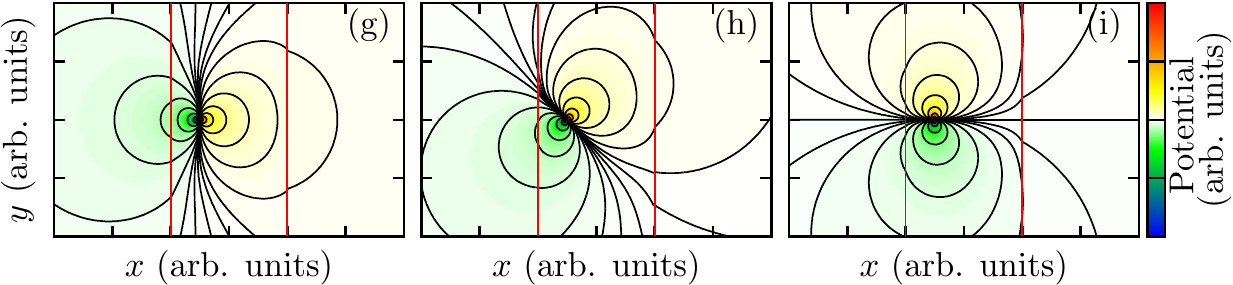}
\caption{Different approximations for the three media problem shown
schematically in Fig.~\ref{179} for different orientations of the dipole.
(a{\textendash}c) include up to
first order reflections, (d{\textendash}f) second order reflections and
(g{\textendash}i) include up to third order reflections. The red lines
indicate the interfaces between different materials: the central material
has an (arbitrary) permittivity of $\varepsilon = 1$,
the material on the left has $\varepsilon = 2$ and the material on the
right has $\varepsilon = 3$. It can be seen that third order reflections
are sufficient to converge the potential for that particular set of
relative values of $\varepsilon$.}
\label{181}
\end{figure*}
Finally the expression of the potential in all
three regions can be
written as
\begin{widetext}
\begin{align}
\phi_{\textbf{p}}^{(1)} \left( \textbf{r} \right) &=
\frac{1}{4 \pi \varepsilon_0 \varepsilon_1} \left\{
\frac{\textbf{p} \cdot \left( \textbf{r} - \textbf{r}_\textbf{p}
\right)}{|\textbf{r} - \textbf{r}_\textbf{p}|^3} + \sum\limits_{n=1}^{\infty}
\left( \frac{\textbf{p}_{(n)}' \cdot \left( \textbf{r} -
\textbf{r}_{\textbf{p}_{(n)}'}\right)}{|\textbf{r} -
\textbf{r}_{\textbf{p}_{(n)}'}|^3} +
\frac{\textbf{p}_{(n)}'' \cdot \left( \textbf{r} -
\textbf{r}_{\textbf{p}_{(n)}''}\right)}{|\textbf{r} -
\textbf{r}_{\textbf{p}_{(n)}''}|^3} \right) \right\}
,
\nonumber \\
\phi_{\textbf{p}}^{(2)} \left( \textbf{r} \right) &=
\frac{1}{4 \pi \varepsilon_0 \varepsilon_2} \left\{
\frac{\textbf{p} \cdot \left( \textbf{r} - \textbf{r}_\textbf{p}
\right)}{|\textbf{r} - \textbf{r}_\textbf{p}|^3} + \sum\limits_{n=1}^{\infty}
\left( \frac{\textbf{p}_{(2n)}' \cdot \left( \textbf{r} -
\textbf{r}_{\textbf{p}_{(2n)}'}\right)}{|\textbf{r} -
\textbf{r}_{\textbf{p}_{(2n)}'}|^3} +
\frac{\textbf{p}_{(2n-1)}'' \cdot \left( \textbf{r} -
\textbf{r}_{\textbf{p}_{(2n-1)}''}\right)}{|\textbf{r} -
\textbf{r}_{\textbf{p}_{(2n-1)}''}|^3} \right) \right\}
\frac{2 \varepsilon_2}{\varepsilon_1 + \varepsilon_2}
,
\nonumber \\
\phi_{\textbf{p}}^{(3)} \left( \textbf{r} \right) &=
\frac{1}{4 \pi \varepsilon_0 \varepsilon_3} \left\{
\frac{\textbf{p} \cdot \left( \textbf{r} - \textbf{r}_\textbf{p}
\right)}{|\textbf{r} - \textbf{r}_\textbf{p}|^3} + \sum\limits_{n=1}^{\infty}
\left( \frac{\textbf{p}_{(2n-1)}' \cdot \left( \textbf{r} -
\textbf{r}_{\textbf{p}_{(2n-1)}'}\right)}{|\textbf{r} -
\textbf{r}_{\textbf{p}_{(2n-1)}'}|^3} +
\frac{\textbf{p}_{(2n)}'' \cdot \left( \textbf{r} -
\textbf{r}_{\textbf{p}_{(2n)}''}\right)}{|\textbf{r} -
\textbf{r}_{\textbf{p}_{(2n)}''}|^3} \right) \right\}
\frac{2 \varepsilon_3}{\varepsilon_1 + \varepsilon_3}
.
\label{180}
\end{align}
\end{widetext}
It is implicit in Eq.~(\ref{180}) that for the calculation of the potential
$\phi_\textbf{p}^{(2)}$ in region (2) only the image dipoles in region (3)
(together with the original dipole) are taken into account, and vice-versa.
Given the form of Eq.~(\ref{180}) it is clear that an exact solution to 
the problem of three media cannot be obtained for a finite number of
terms in the summation. However, approximate solutions can be obtained
whose accuracy will depend mostly on the difference in the values of the
dielectric constants of the different materials.
In Fig.~\ref{181}
we show approximations up to third order reflections,
for different orientations
of the dipole, in the case of three materials for which
$\varepsilon_2 = 2 \varepsilon_1$ and $\varepsilon_3 = 3 \varepsilon_1$.
This is an extreme case in the context of III-V compounds, for which the
differences in $\varepsilon_r$
between materials do not usually go beyond 50\%. For clarity
of interpretation, the potential isolines shown decay as a power of 2,
which allows to visualize the fine effects of the interfaces far from the
dipole. As can be seen,
the second order correction [Figs.~\ref{181}(d{\textendash}f)]
is already very well converged for this extreme case
and we expect first order corrections to be sufficient for the materials of
interest, group-III nitrides in particular.

\subsection{Gaussian smearing of point dipoles}\label{219}

As mentioned in the paper, the potential solution
for a point dipole is an approximation to the potential due to the
dipole moment of a charge distribution.~\cite{jackson_1999}
This approximation is only valid
in the limit when the potential is calculated sufficiently far away from
the charge distribution. How far is ``sufficiently far'' depends on the
particular problem at hand, basically on the value of the dipole and the
volume over which the charge density giving rise to the dipole moment spread
originally. A Gaussian smearing of the dipoles
that are close to the position
where the potential is calculated, is
a straightforward manner to deal with this problem, as the parameters
controlling the smearing can be tuned easily at need. We propose the
implementation of this smearing controlled by two parameters:
\begin{enumerate}
\item $r_\text{smear}$ is the cut-off radius for which all the dipoles that
obey $|\textbf{r}-\textbf{r}_\textbf{p}| < r_\text{smear}$ are smeared,
where $\textbf{r}$ is the position where the potential is calculated and
$\textbf{r}_\textbf{p}$ is the position of the dipole under consideration.
\item $\sigma$ is the standard deviation of the
Gaussian function that produces the smearing. It gives a measure of the
volume over which the dipole is smeared.
\end{enumerate}
Therefore, the expression for the potential
$\phi_\textbf{p} \left( \textbf{r} \right)$ at $\textbf{r}$ due to a
dipole $\textbf{p}$ located at $\textbf{r}_\textbf{p}$ can be rewritten,
in spherical coordinates, as
\begin{widetext}
\begin{align}
\phi_{\textbf{p}} \left( \textbf{r} \right) = 
\begin{cases}
\dfrac{1}{4 \pi \varepsilon_0
\varepsilon_r} \dfrac{\textbf{p} \cdot \left( \textbf{r} - \textbf{r}_\textbf{p}
\right)}{|\textbf{r} - \textbf{r}_\textbf{p}|^3} \quad &\text{for }
|\textbf{r}-\textbf{r}_\textbf{p}| \geq r_\text{smear}
\\
\dfrac{1}{4 \pi \varepsilon_0 \varepsilon_r}
\dfrac{1}{\left( 2 \pi \sigma^2 \right)^\frac{3}{2}}
\int\limits_{0}^{2\pi} \text{d}\varphi
\int\limits_{0}^{\pi} \text{d}\theta
\int\limits_{0}^{\infty} \text{d}\rho
\sin{\theta} \rho^2
e^{-\frac{|\textbf{r}-\textbf{r}_\textbf{p}'|^2}{2 \sigma^2}}
\dfrac{\textbf{p} \cdot \left( \textbf{r} - \textbf{r}_\textbf{p}'
\right)}{|\textbf{r} - \textbf{r}_\textbf{p}'|^3} \quad &\text{for }
|\textbf{r}-\textbf{r}_\textbf{p}| < r_\text{smear}
\end{cases}
\label{182}
\end{align}
\end{widetext}
where $\textbf{r}_\textbf{p}'$ is given by
\begin{align}
\textbf{r}_\textbf{p}' = \textbf{r}_\textbf{p} +
\left[
\rho \sin{\theta} \cos{\varphi},\,
\rho \sin{\theta} \sin{\varphi},\,
\rho \cos{\theta}
\right].
\end{align}
Typically, the integration in $\rho$ can be done up to a certain cutoff
since the value of the integrand will decay rapidly. For example,
our current implementation sets $3.4\sigma$
as the upper limit for the integration, which comprises
a volume that contains about 99\% of the total original dipole moment
$\textbf{p}$.
The extension of Eq.~(\ref{182})
to the case in which different dielectric constants are
present is straightforward and done
in the same way as explained in the paper and Section~\ref{185} of this
Appendix.

\subsection{Computational aspects: method of layers and
application to quantum wells}\label{37}\label{220}

\begin{figure}[t]
\includegraphics{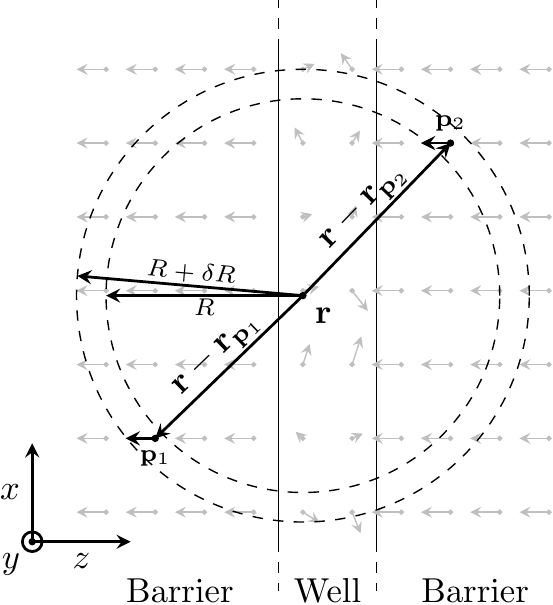}
\caption{Schematic representation of the dipoles present in
a typical nitride QW structure. In a sphere of radius $R$ from $\textbf{r}$
there exist a certain number of dipole pairs for which
\mbox{$\textbf{p}_1 \cdot \left( \textbf{r} - \textbf{r}_{\textbf{p}_1} \right)
= - \textbf{p}_2 \cdot \left( \textbf{r} - \textbf{r}_{\textbf{p}_2} \right)$}
and therefore tend to neutralize each
other (they do not exactly cancel each other due to the image dipole
effect that depends on how far $\textbf{r}$ and $\textbf{r}_\textbf{p}$
are from \textit{each}
interface). For large $R$ this cancellation
effect is bigger as the polarization
is usually constant in the barrier.}
\label{46}
\end{figure}

\begin{figure}[t]
\includegraphics{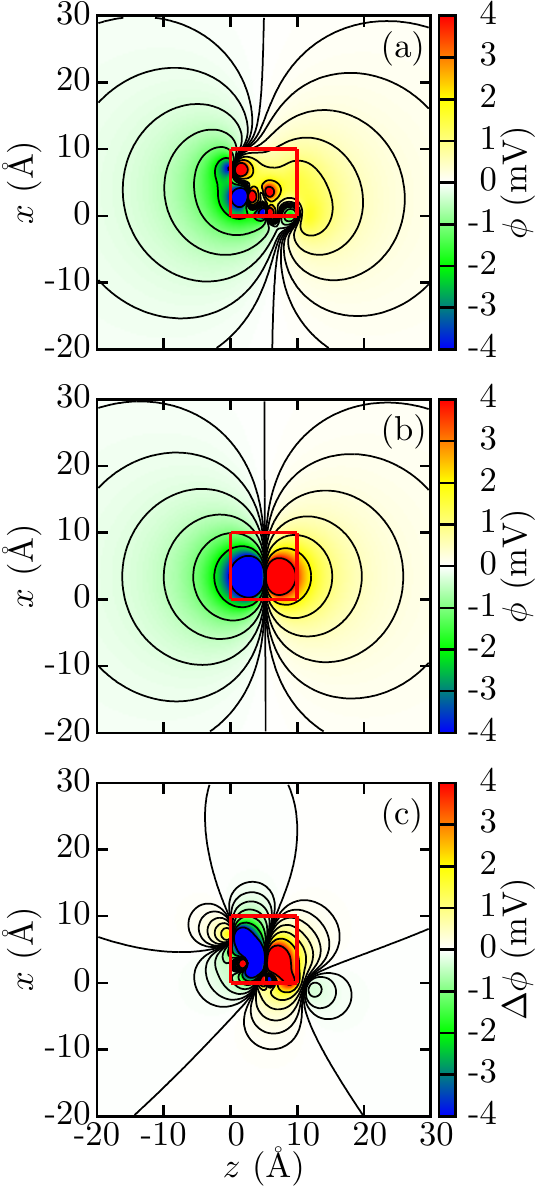}
\caption{Section in the $xz$ plane of:
(a) The polarization potential due
to an ensemble of 10 dipoles randomly placed inside a cube of
side {10~\AA}  with origin at $(0,0,0)$. The boundaries of the cube
within the $xz$ plane is indicated by the red line. The dipoles components are
given by a Gaussian probability distribution with $p_x$ centered at
$10^{-31}$ Cm, $p_y$ and $p_z$ centered at zero, and standard deviation
$5 \times 10^{-32}$ Cm for all three components; (b) The polarization
potential due to a single dipole obtained from the ensemble in (a)
calculated by means of Eq.~(\ref{183}); And (c) difference
between the potentials shown in (b) and (a). Note that the
potential isolines shown follow an exponential behaviour to
exaggerate the results: the lines escaping the plots correspond to zero
and the outer lobe-shaped isolines indicate $\sim 6 \mu\text{V}$.}
\label{184}
\end{figure}

It is clear that when dealing with real size structures, for which the
polarization is sampled at a very elevated number of sites,
the calculation of the potential
$\phi \left( \textbf{r} \right)$ becomes very expensive. In particular,
for each $\textbf{r}$, a summation over \textit{all} the dipoles
present in the system has
to be carried out:
\begin{align}
\phi \left( \textbf{r} \right) = \sum\limits_\textbf{p} \phi_\textbf{p}
\left( \textbf{r} \right).
\label{44}
\end{align}
In a system where the density of dipoles $n_\textbf{p}$
is approximately constant,
for instance one dipole located at each cation site
in Ref.~\onlinecite{caro_2012},
the number of dipoles $\delta N_\textbf{p}$
contributing to Eq.~(\ref{44}) located at distances between
$R$ and $R + \delta R$ from $\textbf{r}$ is proportional to the surface area
of a sphere of radius $R$:
\begin{align}
\delta N_\textbf{p} \propto 4 \pi R^2 n_\textbf{p} \delta R,
\label{45}
\end{align}
where $\delta R$ is an infinitesimal increment in $R$. Because the
contribution to $\phi(\textbf{r})$ from each dipole decreases like
$1/R^2$, as given by Eq.~(\ref{39}), Eq.~(\ref{45}) implies that
the contribution to $\phi (\textbf{r})$ due to the dipoles located at
$\textbf{r}_\textbf{p}$ for which
$R < |\textbf{r} - \textbf{r}_\textbf{p}| < R + \delta R$ is of the
same order of magnitude as the contribution due to dipoles for
which $R' < |\textbf{r} - \textbf{r}_\textbf{p}| < R' + \delta R$,
for any arbitrary $R' > R$. In other words, in principle, the sum in
Eq.~(\ref{44}) does not converge. In practice, for real structures
such as InGaN/GaN QWs, the fact that there is a dot
product involved in the calculation of the potential due to each dipole,
and also that the dipoles in the barrier typically point in the same direction,
give rise to opposite contributions that tend to cancel each other
as $R$ increases,
as schematically shown in Fig.~\ref{46}. In that
case, the sum does converge although rather slowly.
We have implemented
two different methods to speed up the convergence of the sum in
Eq.~(\ref{44}), one of which can be applied to any system, the ``method
of layers''. The other method can be applied to systems where some
assumption can be made about the value of the polarization being constant
in the greatest part of the system, as is the case in QWs. An outline of
these methods is given next.

\subsubsection{Method of layers}

In the same way that a point dipole is an approximation for a charge
density distribution valid far away from the location of the dipole,
it can be shown that a point dipole can be a valid approximation for a
given \textit{ensemble} of neighbouring point dipoles at a
certain distance from the ensemble.
Figure~\ref{184}(a) shows the potential due to an ensemble of $N$ dipoles
$\textbf{p}_i$, of
typical magnitude in nitride QWs,
that are localized in a restricted region in space, each at position
$\textbf{r}_{\textbf{p}_i}$.
This ensemble can be approximated by a single dipole $\textbf{P}$
whose magnitude equals the summation of
all the original dipoles and whose position $\textbf{r}_\textbf{P}$
is given by the weighted average of the dipoles in the ensemble
[Fig.~\ref{184}(b)]:
\begin{align}
\textbf{P} = \sum\limits_{i=1}^N \textbf{p}_i ,
\qquad
\textbf{r}_\textbf{P} = \frac{1}{|\textbf{P}|}
\sum\limits_{i=1}^N \textbf{r}_{\textbf{p}_i} \, |\textbf{p}_i|.
\label{183}
\end{align}
As shown in Fig.~\ref{184}(c), the difference between an ensemble
of dipoles and its correspondent approximation calculated as
in Eq.~(\ref{183}) decays rapidly away from the ensemble.
Applying Eq.~(\ref{183}) recurrently, one can construct,
around the point $\textbf{r}$
where the potential $\phi$ is being
calculated, a system of ``layers''
in which the density of dipoles decreases as one moves
away from $\textbf{r}$.

\subsubsection{Simplification for quantum wells}

A simplification can be made for QW systems, or even
a quantum dot (QD) system, if a constant value for the polarization can be
assumed for the greatest part of the system. Since only \textit{differences}
in polarization are meaningful for the calculation of polarization
potentials, an arbitrary constant shift
of the polarization of the \textit{whole}
system will not have any effect on the calculated value of the polarization
potential. This shift can be chosen in such a way that the
resultant polarization, at least on average,
is zero in the barrier in the case of a QW,
or in the \textit{unstrained} barrier in the case of a
QD.~\footnote{The strain field will not vanish in the barrier
material immediately around a QD, but will effectively be zero
far away from it. See for instance Ref.~\onlinecite{pryor_1998}.}
In that case, all the dipoles arising from that region, once the
discretization described in the paper is made, will
have value zero. Therefore, the dipoles contained within
that region can be left out of the calculation.

\end{document}